\documentclass[a4paper,onecolumn,11pt,accepted=2026-11-16]{quantumarticle}
\pdfoutput=1
\usepackage[utf8]{inputenc}
\usepackage[english]{babel}
\usepackage[T1]{fontenc}
\usepackage{amsmath}
\usepackage{mathtools}
\usepackage{hyperref}
\usepackage[numbers]{natbib}
\usepackage{graphicx} 
\usepackage{multirow}
\usepackage{xcolor}
\usepackage{comment}

\DeclarePairedDelimiter\bra{\langle}{\rvert}
\DeclarePairedDelimiter\ket{\lvert}{\rangle}

\begin{document}

\title{Improved parameter initialization for the (local) unitary cluster Jastrow ansatz}

\author{Wan-Hsuan Lin}
\email{wanhsuanlin@g.ucla.edu}
\affiliation{IBM Quantum, IBM T. J. Watson Research Center, Yorktown Heights, NY 10598, USA}
\affiliation{Computer Science Department, University of California, Los Angeles, CA 90094, USA}

\author{Fangchun Liang}
\affiliation{Center for Computational Life Sciences, Lerner Research Institute, The Cleveland Clinic, Cleveland, Ohio 44106, USA}

\author{Mario Motta}
\affiliation{IBM Quantum, IBM T. J. Watson Research Center, Yorktown Heights, NY 10598, USA}

\author{Haimeng Zhang}
\affiliation{IBM Quantum, IBM T. J. Watson Research Center, Yorktown Heights, NY 10598, USA}

\author{Kenneth M. Merz Jr.}
\affiliation{Center for Computational Life Sciences, Lerner Research Institute, The Cleveland Clinic, Cleveland, Ohio 44106, USA}
\affiliation{Department of Chemistry, Michigan State University, East Lansing, Michigan 48824, United States}

\author{Kevin J. Sung}
\email{kevinsung@ibm.com}
\affiliation{IBM Quantum, IBM T. J. Watson Research Center, Yorktown Heights, NY 10598, USA}

\maketitle

\begin{abstract}
The unitary cluster Jastrow (UCJ) ansatz and its variant known as local UCJ (LUCJ) are promising choices for variational quantum algorithms for chemistry due to their combination of physical motivation and hardware efficiency. The parameters of these ansatzes can be initialized from the output of a coupled cluster, singles and doubles (CCSD) calculation performed on a classical computer. However, truncating the number of repetitions of the ansatz, as well as discarding interactions to accommodate the connectivity constraints of near-term quantum processors, degrade the approximation to CCSD and the resulting energy accuracy. In this work, we propose two methods to improve the parameter initialization. The first method, which is applicable to both expectation value- and sample-based algorithms, uses compressed double factorization of the CCSD amplitudes to improve or recover the CCSD approximation. The second method, which is applicable to sample-based algorithms, uses approximate tensor network simulation to improve the quality of samples produced by the ansatz circuit. Both methods perform classical preprocessing to construct the initial circuit parameters and can be followed by parameter optimization using quantum measurements, although we do not perform such optimization in this work. We validate our methods using exact state vector simulation on systems of up to 52 qubits, as well as experiments on superconducting quantum processors using up to 65 qubits. Our results indicate that our methods can significantly improve the output of both expectation value- and sample-based quantum algorithms.
\end{abstract}

\section{Introduction}

Quantum computing holds significant promise for quantum chemistry applications, particularly in cases with strongly correlated electronic states~\cite{mcardle2020quantum,motta2022emerging}. Variational quantum algorithms using parameterized circuit ansatzes are a common approach for estimating ground state energies on pre-fault-tolerant quantum processors. Compared to alternative approaches that do not use parameterized circuits, such as those based on Krylov subspaces~\cite{yoshioka2025krylov,yu2025quantumcentricalgorithmsamplebasedkrylov}, variational quantum algorithms provide flexibility in circuit depth and structure. These algorithms can broadly be classified into expectation value- and sample-based methods, depending on whether they require quantum estimation of expectation values. 
The variational quantum eigensolver (VQE)~\cite{peruzzo2014variational} and quantum-selected configuration interaction (QSCI)~\cite{kanno2026qsci} serve as paradigmatic examples of expectation value- and sample-based algorithms, respectively.

A key challenge in designing variational quantum algorithms is selecting an appropriate circuit ansatz. Physics-inspired ansatzes, such as unitary coupled cluster (UCC) and its variants~\cite{anand2022uccsd}, have the favorable property that their parameters can be initialized from classical approximations. However, implementing these ansatzes on near-term quantum processors is infeasible due to the enormous number of gates required~\cite{motta2020quantum_simulation_of_electronic_structure_with_transcorrelated_hamiltonian}.
On the other hand, hardware-efficient ansatzes are designed to minimize circuit depth and adapt to the connectivity of devices~\cite{kandala2017hardwareefficient}, but they lack physical motivation, break physical symmetries, and can be challenging to initialize and optimize~\cite{larocca2025barren,Dcunha2023hardware_efficient_ansatze}.

Recently, the unitary cluster Jastrow (UCJ) ansatz~\cite{matsuzawa2020ucj} has been proposed as an alternative that adopts the repetition structure of a UCC operator with a low-rank decomposition while improving hardware compatibility by truncating the number of repetitions to reduce circuit depth.
Its variant, local unitary cluster Jastrow (LUCJ) ansatz~\cite{motta2023lucj}, has attracted growing attention due to its combination of physical motivation and hardware efficiency~\cite{blunt2025quantum,liepuoniute2025quantumcentric,kaliakin2025implicit_solvent,robledo2025sqd,shajan2025toward}.
The LUCJ ansatz adapts UCJ for hardware with limited connectivity by discarding interactions that would otherwise incur a SWAP overhead. 
Like the UCJ ansatz, LUCJ can be initialized from $t$-amplitudes obtained from a coupled cluster, singles and doubles (CCSD)~\cite{bartlett2007coupled} calculation, though some accommodation must be made for the connectivity constraints. Prior works have simply discarded interactions incompatible with those constraints, but this approach can severely degrade the quality of the approximation to CCSD. Even in the context of UCJ without connectivity constraints, truncating the number of ansatz repetitions also degrades the quality of the approximation.

In this work, we propose two complementary methods to improve parameter initialization for the LUCJ ansatz. The first method, which is applicable to both expectation value- and sample-based algorithms, uses compressed double factorization~\cite{Cohn_2021_compressed_df} to better approximate the original $t$-amplitudes from CCSD. The second method, which is applicable to sample-based algorithms, uses approximate tensor network simulation to improve the quality of samples produced by the ansatz circuit, and is a type of surrogate optimization~\cite{2023ClassicalSurrogates,hirsbrunner2024beyond,2025PNAS}. While previous work has explored surrogate optimization with tensor networks in the context of expectation value-based algorithms~\cite{rudolph2023synergistic,khan2023preoptimizing}, we show that it can also be effective for sample-based algorithms.

Our study evaluates the quality of the states prepared with these classically initialized parameters. The compressed double factorization optimizes a classical amplitude-fitting objective, and the tensor network method further refines the parameters by minimizing an approximate QSCI energy evaluated entirely on a classical computer. These procedures are part of the initialization; we do not subsequently optimize the parameters using quantum measurements, nor do we run a VQE-style optimization loop. Whether these initializations accelerate subsequent quantum optimization, avoid local minima, or improve trainability is left to future work.

Our methods are distinct from other parameter initialization strategies for variational quantum algorithms. Parameter-transfer approaches reuse an ansatz and optimized parameters from a smaller, related problem to initialize a larger one~\cite{liu2023mitigating}. In contrast, our methods do not require a previously optimized instance, but rather start from a classical CCSD calculation. Other recent works have established theoretical conditions under which initialization in certain regions of parameter space avoids exponentially vanishing gradients~\cite{park2024hamiltonian,mhiri2025unifyingaccountwarmstart}. Our methods are problem-informed heuristics for constructing initial parameters that retain information from CCSD, and we do not provide guarantees regarding gradient scaling or trainability. Establishing whether our initialization heuristics place UCJ or LUCJ circuits in trainable regions is an interesting direction for future work.

We validate our methods using exact state vector simulation on systems of up to 52 qubits, as well as experiments on superconducting quantum processors using up to 65 qubits. Our results indicate that our methods can significantly improve the output of both expectation value- and sample-based quantum algorithms. We have contributed an implementation of the compressed double factorization to the open-source software library \textsf{ffsim}~\cite{ffsim}.

\section{Background}

\subsection{Expectation value- and sample-based algorithms}

Quantum algorithms that use variational circuit ansatzes can broadly be classified into expectation value- and sample-based algorithms based on whether they require estimation of the quantity $\bra{\Psi} H \ket{\Psi}$ on the quantum computer, where $\ket{\Psi}$ is the ansatz state and $H$ is an observable. The canonical expectation value-based algorithm is the variational quantum eigensolver (VQE)~\cite{peruzzo2014variational}. In VQE, $H$ is a Hamiltonian and a classical optimization is performed to find ansatz parameters that minimize $\bra{\Psi} H \ket{\Psi}$. In the rest of this paper, we refer to $\bra{\Psi} H \ket{\Psi}$ as the Hamiltonian expectation value of the ansatz state $\ket{\Psi}$, labeled ``Expectation value'' in the figures. We evaluate this quantity at the initialized parameters by classical state vector simulation, without a VQE optimization loop. It is used as a metric of initialization quality and does not imply a converged VQE calculation.

Due to the large sampling overhead involved in obtaining a precise estimate of the Hamiltonian expectation value~\cite{wecker2015progress}, there has been interest in sample-based algorithms that do not require estimating the Hamiltonian expectation value on the quantum computer. Quantum-selected configuration interaction (QSCI)~\cite{kanno2026qsci} is a paradigmatic example of a sample-based algorithm. In QSCI, the ansatz state is sampled to produce configurations, and the Hamiltonian is projected and diagonalized inside a subspace formed by these configurations. In the rest of this paper, we refer to the estimate of the ground state energy obtained by this procedure as the ``QSCI energy'' of the ansatz state $\ket{\Psi}$. Note that the QSCI energy is actually a random variable because it depends on the measurement outcomes, and its precise definition also depends on experimental parameters such as how many configurations to include in the diagonalization subspace. Like in VQE, the QSCI energy can be minimized using a classical optimization.

An extension of QSCI known as sample-based quantum diagonalization (SQD)~\cite{robledo2025sqd} applies an error mitigation procedure, called configuration recovery, to samples that violate known symmetries of the ansatz state. Configuration recovery greatly improves results on noisy quantum hardware and has led to the rapid adoption of SQD for hardware demonstrations~\cite{liepuoniute2025quantumcentric,kaliakin2025implicit_solvent,shajan2025toward}. For our hardware results that incorporate configuration recovery, we refer to the resulting estimate of the ground state energy as the ``SQD energy.''

\subsection{The LUCJ ansatz}

The LUCJ ansatz~\cite{motta2023lucj} is a specialized form of the more general unitary cluster Jastrow (UCJ) ansatz~\cite{matsuzawa2020ucj}, which has the form
\begin{align}
  \lvert \Psi \rangle = \prod_{\mu = 0}^{L - 1} \mathcal{U}_\mu e^{i \mathcal{J}_\mu} \mathcal{U}_\mu^\dagger \lvert \Phi_0 \rangle,
\end{align}
where $\lvert \Phi_0 \rangle$ is a reference state, often taken as the Hartree-Fock state, each $\mathcal{U}_\mu$ is an orbital rotation, and each $\mathcal{J}_\mu$ is a diagonal Coulomb operator of the form
\begin{align}
    \mathcal{J} = \frac12\sum_{ij,\sigma \tau} J^{\sigma \tau}_{ij} n_{i,\sigma} n_{j,\tau},
\end{align}
where $n_{i,\sigma} = c^\dagger_{i,\sigma} c_{i,\sigma}$ is the occupation number operator for the fermionic mode with index $i$ and spin $\sigma$. Here, $J^{\sigma \tau}_{ij}$ denotes the entry in row $i$ and column $j$ of the real symmetric matrix $J^{\sigma \tau}$. In this work, we assume that $J^{\alpha\alpha} = J^{\beta\beta}$ and $J^{\alpha\beta} = J^{\beta\alpha}$ in order to treat spin up and spin down symmetrically, which is desirable for closed-shell molecular systems. As a result, each diagonal Coulomb operator is described by two matrices, $J^{\alpha\alpha}$ and $J^{\alpha\beta}$. However, our results can readily be extended to the case where the spin up and spin down are not treated symmetrically.

Implementing the $e^{i \mathcal{J}_\mu}$ term of the UCJ ansatz requires either all-to-all connectivity or the use of a fermionic swap network, 
making it challenging for noisy pre-fault-tolerant quantum processors that have limited connectivity. 
The idea of the local UCJ (LUCJ) ansatz is to impose sparsity constraints on the $J^{\alpha\alpha}$ and $J^{\alpha\beta}$ matrices such that the resulting circuit can be implemented with a limited and controllable number of SWAP operations (possibly none) on hardware with restricted connectivity.
The constraints are specified by a list of indices indicating which matrix entries are allowed to be nonzero, and these indices can be interpreted as pairs of orbitals that are allowed to interact.

\subsection{Parameter initialization from CCSD}

This section explains how the parameters of the LUCJ ansatz have so far been initialized from the results of a coupled cluster, singles and doubles (CCSD) calculation.

\subsubsection{CCSD and UCCSD}
CCSD is a computational method that approximates the ground state wavefunction with an exponential ansatz~\cite{bartlett2007coupled},
\begin{align}
\lvert \Psi_\text{CCSD} \rangle = e^{T}\lvert \Phi_0 \rangle,
\end{align}
where
\begin{align}
T = T_1 + T_2
\end{align}
with
\begin{align}
T_1 = \sum_{ia} t_{ia} \hat{c}^\dagger_a \hat{c}^{\phantom{\dagger}}_i, \quad
T_2 = \sum_{ijab} t_{ijab} \hat{c}^\dagger_a \hat{c}^\dagger_b \hat{c}^{\phantom{\dagger}}_j \hat{c}^{\phantom{\dagger}}_i.
\label{eq:t-amplitudes}
\end{align}
The ansatz is parameterized by the numbers $t_{ia}$ and $t_{ijab}$, called $t$-amplitudes (separately, $t_1$- and $t_2$-amplitudes), 
which can be efficiently solved for on a classical computer. Here, $i$ and $j$ index occupied orbitals and $a$ and $b$ index virtual orbitals.

Unitary CCSD (UCCSD) is a variant of CCSD that uses a similar exponential ansatz, but with an operator that is guaranteed to be unitary~\cite{anand2022uccsd}:
\begin{align}
\lvert \Psi_\text{UCCSD} \rangle = e^{T - T^\dagger}\lvert \Phi_0 \rangle.
\end{align}
While UCCSD is challenging to implement on a classical computer, it can be implemented efficiently on a quantum computer~\cite{harsha2018on_the_difference}.
Like CCSD, UCCSD is also parameterized by the $t$-amplitudes $t_{ia}$ and $t_{ijab}$. 
If the $t$-amplitudes from CCSD are plugged into UCCSD, and $\lvert \Phi_0 \rangle$ is the Hartree-Fock state, 
then because $T^\dagger \lvert \Phi_0 \rangle = 0$, 
the resulting ansatz states match up to first order in $T$ in the Taylor series of the exponential operators. 
Therefore, the $t$-amplitudes obtained from a CCSD calculation may serve as reasonable initial parameters for UCCSD.

It is useful to distinguish the purpose of this quantum circuit initialization from that of established post-CCSD classical methods. For predominantly single-reference systems, methods such as CCSD(T) generally improve on CCSD and are likely to be preferable whenever they are computationally affordable and the single-reference approximation remains reliable~\cite{raghavachari1989afifthorder,bartlett2007coupled}. Our compressed double factorization method should not itself be viewed as a post-CCSD correlation correction, because it targets CCSD amplitudes and inherits limitations of that starting approximation. Rather, it supplies physically motivated parameters for VQE and QSCI/SQD calculations, after which variational quantum or tensor-network optimization can move beyond the initial classical description. Determining when this approach provides a favorable cost-accuracy tradeoff relative to CCSD(T) and other high-accuracy classical methods remains an important subject for future benchmarking.

\subsubsection{From UCCSD to (L)UCJ}

The UCJ ansatz can be related to the UCCSD ansatz via a double-factorized representation of the $T_2$ operator~\cite{matsuzawa2020ucj}. 
We can focus on the $T_2$ operator because the $T_1$ operator can be separated by invoking a Trotter approximation:
\begin{align}
e^{T - T^\dagger} = e^{T_1 - T_1^\dagger + T_2 - T_2^\dagger} \approx e^{T_1 - T_1^\dagger} e^{T_2 - T_2^\dagger}.
\end{align}
Because $T_1$ contains only quadratic terms, $e^{T_1 - T_1^\dagger}$ is an orbital rotation, and can be implemented efficiently on a quantum computer. 
To approximate $e^{T_2 - T_2^\dagger}$, we represent $T_2 - T_2^\dagger$ in a double-factorized form~\cite{motta2021lowrank},
\begin{align}
T_2 - T_2^\dagger = i\sum_{\mu=0}^{L - 1} \mathcal{U}_\mu \mathcal{J}_\mu \mathcal{U}_\mu^\dagger.
\label{eq:uccsd_df}
\end{align}
The $\mathcal{U}_\mu$ are orbital rotations, and the $\mathcal{J}_\mu$ are diagonal Coulomb operators of the form $\mathcal{J}=\frac12\sum_{\sigma \tau, ij} J_{ij}^{\sigma \tau} n_{\sigma, i} n_{\tau, j}$,
where $n_{\sigma, i}$ is the occupation number operator and $J_{ij}$ is a real symmetric matrix. The number of terms $L$ scales linearly with the number of orbitals with increasing basis size, but quadratically with the number of orbitals with increasing molecular size~\cite{motta2021lowrank}.

Finally, to obtain an operator in UCJ form, we invoke a Trotter approximation:
\begin{align}
e^{T_2 - T_2^\dagger} = e^{i\sum_{\mu=0}^{L - 1} \mathcal{U}_\mu \mathcal{J}_\mu \mathcal{U}_\mu^\dagger} \approx \prod_{\mu=0}^{L - 1} \mathcal{U}_\mu e^{i\mathcal{J}_\mu} \mathcal{U}_\mu^\dagger.
\label{eq:uccsd_to_ucj}
\end{align}
Thus, the $t$-amplitudes obtained from a CCSD calculation may also yield an initial guess for UCJ via double factorization of the $t_2$-amplitudes. If a final orbital rotation is included in the ansatz~\cite{moreno2023enhancing}, it can be initialized from the $t_1$-amplitudes. An LUCJ ansatz can be initialized by first constructing the UCJ ansatz, and then discarding the terms in the diagonal Coulomb operators that are unavailable due to the locality constraints.

\begin{figure}
    \centering
    \includegraphics[width=\linewidth]{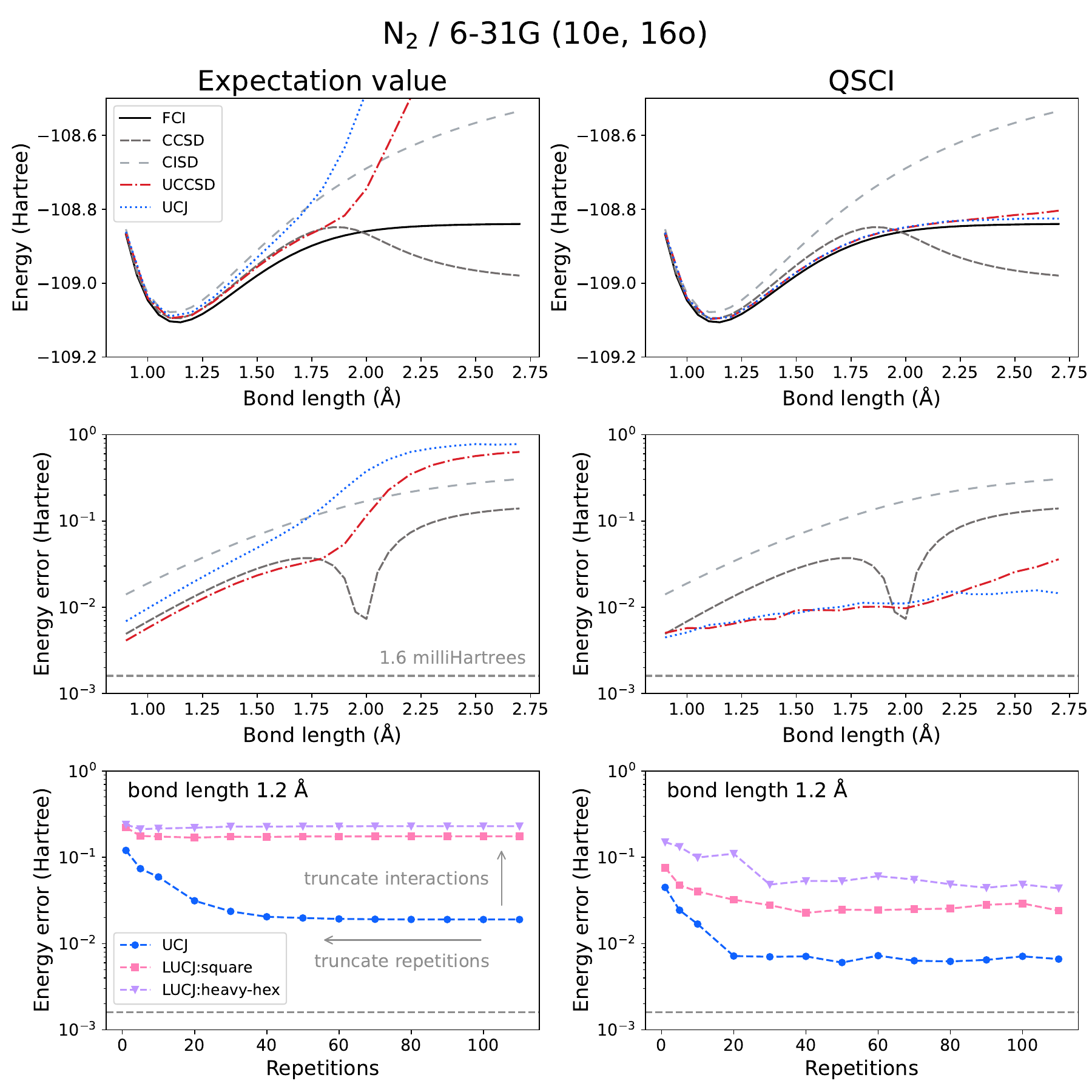}
    \caption{Potential energy curves and the effect of ansatz truncation. Data for N\textsubscript{2} in a (10e,~16o) active space derived from the 6-31G basis set. The left column shows the Hamiltonian expectation value evaluated by classical state vector simulation at the initialized parameters, and the right column shows the ``QSCI energy'' obtained by sampling configurations from the wavefunction and using them to run QSCI. No further optimization of the circuit parameters is performed for the data shown here. For the relatively small N\textsubscript{2}/6-31G system, the variance of the QSCI energy was smaller than numerical precision with our experimental parameters (see Section~\ref{sec:computational_details}), so error bars are not needed. \textbf{Top row}: Potential energy curves obtained by taking the CCSD $t$-amplitudes and plugging them into UCCSD and UCJ. Energy values for full configuration interaction (FCI), CCSD, and CISD are shown for reference. \textbf{Middle row}: Absolute energy error, computed as deviation from FCI. The horizontal line indicates 1.6 milliHartrees, a commonly cited accuracy target for chemistry. Note that $1 \text{ Hartree} \approx 4.36 \times 10^{-18} \text{J}$. \textbf{Bottom row}: Energy error as a function of the number of repetitions retained in the ansatz, for bond length 1.2 Å. Data is shown for the UCJ ansatz, as well as variants of the LUCJ ansatz designed for square lattice and heavy-hex lattice qubit connectivity.}
    \label{fig:energy_comp_and_truncation}
\end{figure}

The top row of Figure~\ref{fig:energy_comp_and_truncation} shows potential energy curves obtained from Hamiltonian expectation values and QSCI for the UCCSD and UCJ ansatzes initialized from CCSD amplitudes. The data is for N\textsubscript{2} in a (10e,~16o) active space derived from the 6-31G basis set. For reference, the CCSD and CISD energies are also shown. The middle row shows the energy error, computed as the deviation from a full configuration interaction (FCI) reference value. For the data in this figure, the Hamiltonian expectation values and QSCI energies are evaluated for the states initialized directly from CCSD amplitudes, with no further optimization of the circuit parameters. The classical initialization improvements introduced in Section~\ref{sec:methods} are evaluated separately below.

The horizontal line at 1.6 milliHartrees in Figure~\ref{fig:energy_comp_and_truncation} should be interpreted as a stringent reference threshold for the absolute energy error, rather than as a direct demonstration of ``chemical accuracy.'' Chemical accuracy conventionally refers to 1.6 milliHartree agreement between computed and experimental energy differences, whereas the errors reported here are differences between approximate and reference total energies within a fixed basis set and active space. Accurately reproducing the exact total electronic energies of the states entering a chemically relevant energy difference is sufficient for achieving chemical accuracy for that quantity, but it may not be necessary because errors in the individual energies may cancel when the energy difference is computed.

The total energy errors in this work generally remain above the 1.6 milliHartree threshold for several reasons. CCSD and CISD do not reach this threshold due to their inherent approximations. Both methods account for only single and double excitations, but for the systems considered here, higher-order excitations have significant contributions to the total energy. Parameter initialization from CCSD or CISD amplitudes inherits these limitations. While a VQE-style quantum optimization of the circuit parameters might bring the errors of UCCSD or (L)UCJ below the 1.6 milliHartree threshold, such optimization is beyond the scope of this work.

\subsubsection{Parameter initialization from CISD}

So far, we have used $t_2$ amplitudes from CCSD calculations to parametrize (L)UCJ quantum circuits. However, CCSD may be inaccurate or fail to converge if the ground-state wavefunction has pronounced multireference character. In such situations, one may parametrize (L)UCJ quantum circuits based on a different wavefunction $\Psi$, e.g. the output of a configuration interaction, singles and doubles (CISD)~\cite{sherrill1999ci} calculation. This wavefunction can be written, without loss of generality, through a linear configuration interaction (CI) expansion as
\begin{equation}
\begin{split}
| \Psi \rangle &= c_0 | \Phi_0 \rangle + \sum_{ia} c^{(1)}_{ia} \hat{c}^\dagger_a \hat{c}^{\phantom{\dagger}}_i | \Phi_0 \rangle + \sum_{aibj} c^{(2)}_{ijab} \hat{c}^\dagger_a \hat{c}^\dagger_b \hat{c}^{\phantom{\dagger}}_j \hat{c}^{\phantom{\dagger}}_i | \Phi_0 \rangle + \dots \\
&= c_0 \left[ | \Phi_0 \rangle + \hat{C}_1 | \Phi_0 \rangle + 
\hat{C}_2 | \Phi_0 \rangle + \dots \right]
\end{split}
\end{equation}
The coefficients defining the CI operators $\hat{C}_1$ and $\hat{C}_2$, which correspond to single and double excitations, may be transformed into CC amplitudes based on the equivalence between the exponential CC ansatz and the linear CI expansion at the basis of methods like tailored CC~\cite{hino2006tailored,morchen2020tailored,scheurer2024tailored}. This equivalence gives the following relation between cluster and CI operators,
\begin{equation}
\hat{C}_1 = \hat{T}_1
\; , \;
\hat{C}_2 = \hat{T}_2 + \frac{1}{2} \hat{T}_1^2 
\; ,
\end{equation}
which in turn allows us to convert the amplitudes $c^{(1)}$ and $c^{(2)}$ into their corresponding CC counterparts as
\begin{equation}
t_{1,ia} = \frac{ c^{(1)}_{ia} }{c_0}
\;,\;
t_{2,ijab} = \frac{ c^{(2)}_{ijab} }{c_0} - \frac{t_{1,ia} t_{1,jb}}{2}
\;.
\end{equation}
The conversion yields amplitudes $t_{1,ia}$ and $t_{2,ijab}$ that one may use to parametrize (L)UCJ quantum circuits in situations where CCSD does not converge and/or is inaccurate.

\subsection{Truncation error}

Initializing the UCJ ansatz from CCSD $t$-amplitudes may yield a reasonable approximation to UCCSD. However, truncating the number of repetitions is often necessary for near-term noisy quantum processors because of the limited number of gates that can be reliably executed.
Truncating the ansatz to a smaller number of repetitions, compared to the full number of terms in the double factorization of the $t_2$-amplitudes, degrades the quality of the approximation for both UCJ and LUCJ ansatzes.
Additionally, discarding interactions to obtain an LUCJ ansatz further degrades the approximation.  

The bottom row of Fig.~\ref{fig:energy_comp_and_truncation} illustrates the effect of ansatz truncation on the Hamiltonian expectation value and QSCI energy. Truncating repetitions from the UCJ ansatz has little effect when the number of repetitions is high, because later repetitions correspond to terms of smaller magnitude in the double factorization of the $t_2$-amplitudes.
However, truncating interactions to go from the UCJ to the LUCJ ansatz immediately causes the error to become large, even close to the full number of repetitions. 
As seen in the figure, the interactions omitted from the LUCJ ansatz are crucial for maintaining the approximation to CCSD, at least with the naive truncation strategy.

\section{Methods}
\label{sec:methods}

\subsection{Recovering the CCSD approximation by compressed double factorization}
\label{sec:method:subsec:compressed_double_factorization}

In this section, we describe a method to improve the parameter initialization of the UCJ and LUCJ ansatzes by using a compressed double factorization of the $t_2$ amplitudes from CCSD.

\subsubsection{Compressed double factorization of \texorpdfstring{$t_2$}{t\_2} amplitudes}

The double factorization of the $t_2$ amplitudes (see Eq.~\ref{eq:t-amplitudes}) returns tensors $U^{(\mu)}_{ij}$ and $J^{(\mu)}_{ij}$ such that~\cite{motta2021lowrank,Cohn_2021_compressed_df}
\begin{align}
t_{ijab} = i \sum_{\mu=0}^{L - 1} \sum_{p,q=0}^{N-1}
    J^{(\mu)}_{pq}
    U^{(\mu)}_{ap} U^{(\mu)*}_{ip} U^{(\mu)}_{bq} U^{(\mu)*}_{jq}.
    \label{eq:double_factorization_t2}
\end{align}
Here, $N$ is the number of spatial orbitals, each $J^{(\mu)}$ is a real symmetric matrix representing a diagonal Coulomb operator, each $U^{(\mu)}$ is a unitary matrix representing an orbital rotation, $i$ and $j$ run over occupied orbitals, and $a$ and $b$ run over virtual orbitals. If the number of terms in the sum is truncated, or if some diagonal Coulomb interactions are discarded by zeroing out the corresponding entries of the $J^{(\mu)}$ matrices (as for the LUCJ ansatz), then the equality no longer holds. However, we can optimize the remaining orbital rotations and diagonal Coulomb operators from the truncated double factorization, aiming to recover $t_2$ amplitudes that match the original ones more closely.
This procedure is referred to as a compressed double factorization because it compresses the information from a full double factorization into a smaller number of terms or interactions~\cite{Cohn_2021_compressed_df}.

Given the $t_2$ amplitudes from CCSD, we apply double factorization to express them using orbital rotations $U^{(\mu)}$ and diagonal Coulomb matrices $J^{(\mu)}$ as in Eq.~\eqref{eq:double_factorization_t2}.
We then retain the $L$ largest terms, as measured by the Frobenius norm of the diagonal Coulomb matrix, resulting in $L$ orbital rotations and $L$ diagonal Coulomb operators.  The parameter $L$ is user-defined and specifies how many terms are preserved after truncation.

In the subsequent discussion, we denote the original orbital rotations and diagonal Coulomb matrices by $U$ and $J$, respectively, 
and use $\bar{U}$ and $\bar{J}$ to represent their compressed counterparts.
Note that given any two tensors $\bar{U}^{(\mu)}_{ij}$ and $\bar{J}^{(\mu)}_{ij}$, 
we can plug them into the expression to obtain some $t_2$ amplitudes $\bar{t}_{ijab}$. 
Therefore, the compressed double factorization attempts to find tensors that minimize the least-squares objective function
\begin{align}
\frac12 \sum_{ijab} \lvert \bar{t}_{ijab} - t_{ijab} \rvert ^2,
\label{eq:objective}
\end{align}
where $\bar{t}_{ijab}$ are the amplitudes reconstructed by the compressed operator,
and $t_{ijab}$ are the original $t_2$ amplitudes. The minimization can be performed using a standard gradient-based optimizer; we use L-BFGS-B~\cite{byrd1995lbfgsb}.

In addition to addressing the performance loss from truncating the number of ansatz repetitions, we also improve upon the degradation introduced by limited connectivity.
The connectivity constraints on $J^{\alpha \alpha}$ and $J^{\alpha \beta}$ are described by the index sets $S_{\alpha \alpha}$ and $S_{\alpha \beta}$~\cite{motta2023lucj}. For example, for square lattice connectivity, the sets may be chosen as
\begin{equation}
\begin{aligned}
S_{\alpha \alpha} &= \{(p, p+1) \; , \; p = 0, \ldots, N-2\}, \\
S_{\alpha \beta} &= \{(p, p),\ \; p = 0, \ldots, N-1\}.
\end{aligned}
\end{equation}
To incorporate these constraints into the optimization, we force matrix entries with indices outside these sets to be zero. In other words, $J^{(\mu)}_{pq}$ is allowed to be nonzero only if $(p, q) \in S_{\alpha \alpha} \cup S_{\alpha \beta}$. This strategy is simple to implement because the expression~(\ref{eq:double_factorization_t2}) for the $t_2$ amplitudes does not distinguish separate $J^{(\mu)}$ matrices for same-spin and opposite-spin interactions. However, because $S_{\alpha \alpha}$ and $S_{\alpha \beta}$ can differ from each other, some interactions still need to be discarded when constructing the LUCJ ansatz, but fewer than if no constraints were considered at all.

\begin{figure}
    \centering
    \includegraphics[width=\linewidth]{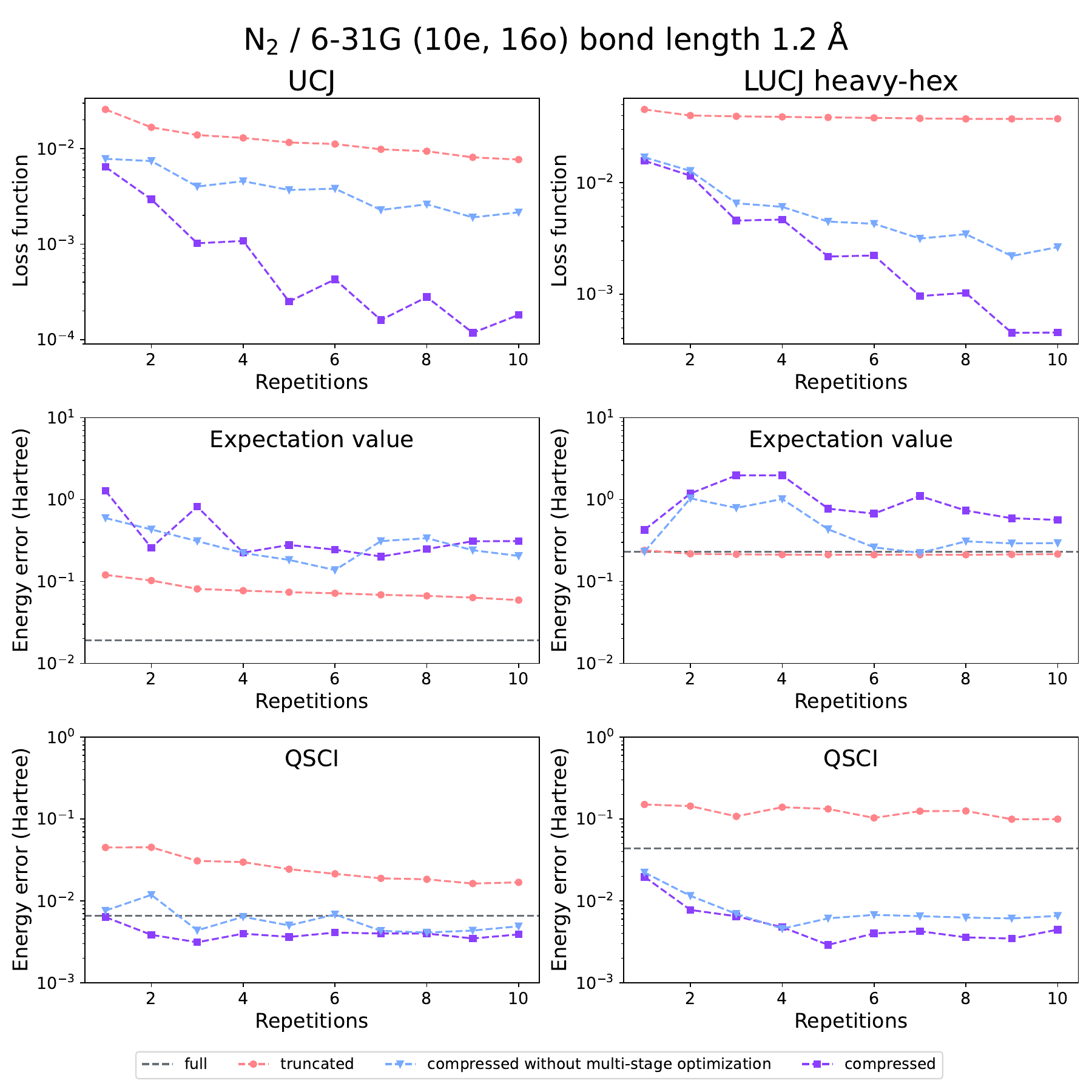}
    \caption{Double-factorization loss function and multi-stage optimization. Data for the (10e,~16o) N\textsubscript{2} system at bond length 1.2 Å. The left column shows data for the UCJ ansatz, and the right column shows data for the LUCJ ansatz with heavy-hex connectivity. \textbf{Top row}: Double-factorization final loss function value as a function of the number of ansatz repetitions, for naive truncation, compressed double-factorization, and compressed double-factorization without multi-stage optimization. \textbf{Middle row}: Error in the Hamiltonian expectation value, computed by classical state vector simulation at the initialized parameters. Here, compressed double-factorization does not improve the Hamiltonian expectation value due to Trotter error, but this can be fixed by adding a regularization term to the loss function (see Fig.~\ref{fig:regularization}). The horizontal line indicates the energy for the ansatz with the full untruncated number of repetitions. \textbf{Bottom row}: QSCI energy error. Compressed double-factorization does consistently improve the QSCI energy.}
    \label{fig:t2_loss}
\end{figure}

\subsubsection{Multi-stage optimization}

A naive approach to optimization would be to truncate the operators down to the target number of terms and then optimize their coefficients. However, the truncated operator may not be a good starting point for optimization as the number of discarded terms increases, and the optimization can converge to a local minimum. To mitigate this issue, we adopt a multi-stage optimization framework for initializing the parameters of the optimization. Instead of directly truncating to the target number of terms, the process iteratively removes some terms, optimizing the parameters at each step, until the target number of terms remains. Because a smaller number of terms is removed at each step, the remaining terms give a better starting point for the optimization.

The first row of Fig.~\ref{fig:t2_loss} shows that multi-stage optimization is effective at reducing the loss function, compared to optimization without the multi-stage procedure. The second row shows that a reduced loss function does not necessarily lead to an improvement in the Hamiltonian expectation value. On the other hand, the third row shows that a reduced loss function does generally lead to an improvement in the QSCI energy.

The reason that a reduced loss function does not necessarily improve the Hamiltonian expectation value is due to the Trotter error in Eq.~\ref{eq:uccsd_to_ucj}. The Trotter error is exacerbated if the norms of the diagonal Coulomb operators become large. To mitigate this issue, we can add a regularization term to the loss function, as described in the next section.

\subsubsection{Operator norm regularization}

To mitigate Trotter error due to increased operator norms, we can add a term to the least-squares objective function that penalizes deviations between the norms of the original and compressed Coulomb matrices.
The objective function with the regularization term is
\begin{align}
\frac12 \sum_{ijab} \lvert \bar{t}_{ijab} - t_{ijab} \rvert ^2 +
\lambda \left\lvert \sum_{\mu} \| \bar{J}^{(\mu)} \|_\text{F}^2 - \sum_{\mu} \| J^{(\mu)} \|_\text{F}^2 \right\rvert,
\end{align}
where $\|\cdot\|_\text{F}$ denotes the Frobenius matrix norm, 
and $\lambda$ is a regularization parameter that controls how strongly to weight the regularization term.

\begin{figure}
    \centering
    \includegraphics[width=\linewidth]{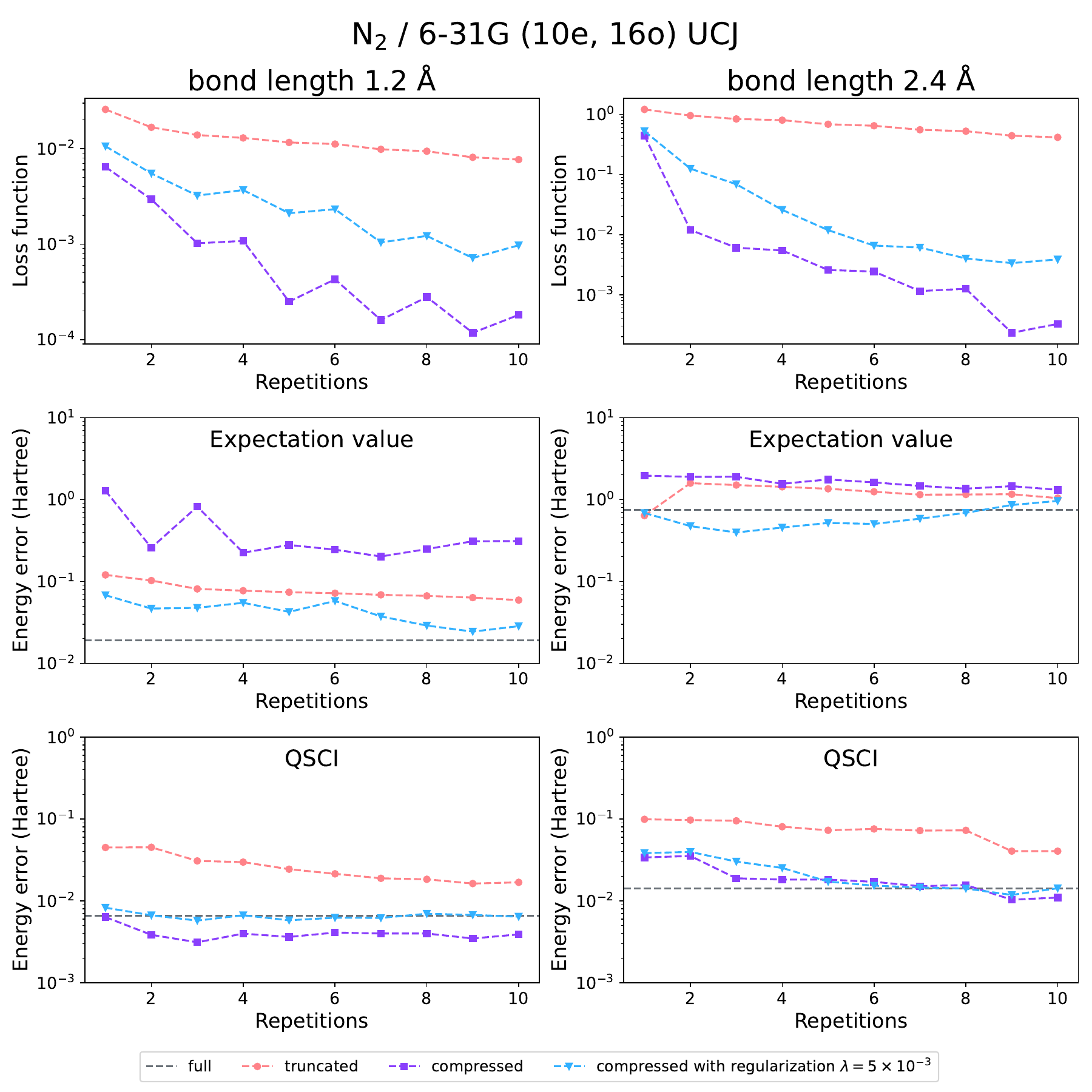}
    \caption{Effect of adding an operator norm regularization term to the loss function. Data for the (10e,~16o) N\textsubscript{2} system with the UCJ ansatz. The left column shows data for bond length 1.2 Å, and the right column shows data for bond length 2.4 Å. \textbf{Top row}: Double-factorization final loss function value as a function of the number of ansatz repetitions, for naive truncation, compressed double-factorization, and compressed double-factorization with regularization. \textbf{Middle row}: Error in the Hamiltonian expectation value, computed by classical state vector simulation at the initialized parameters. With regularization, compressed double-factorization usually improves the Hamiltonian expectation value. The horizontal line indicates the energy for the ansatz with the full untruncated number of repetitions. \textbf{Bottom row}: QSCI energy error. For QSCI, regularization is not needed and can actually increase the energy error.}
    \label{fig:regularization}
\end{figure}

\begin{figure}
    \centering
    \includegraphics[width=\linewidth]{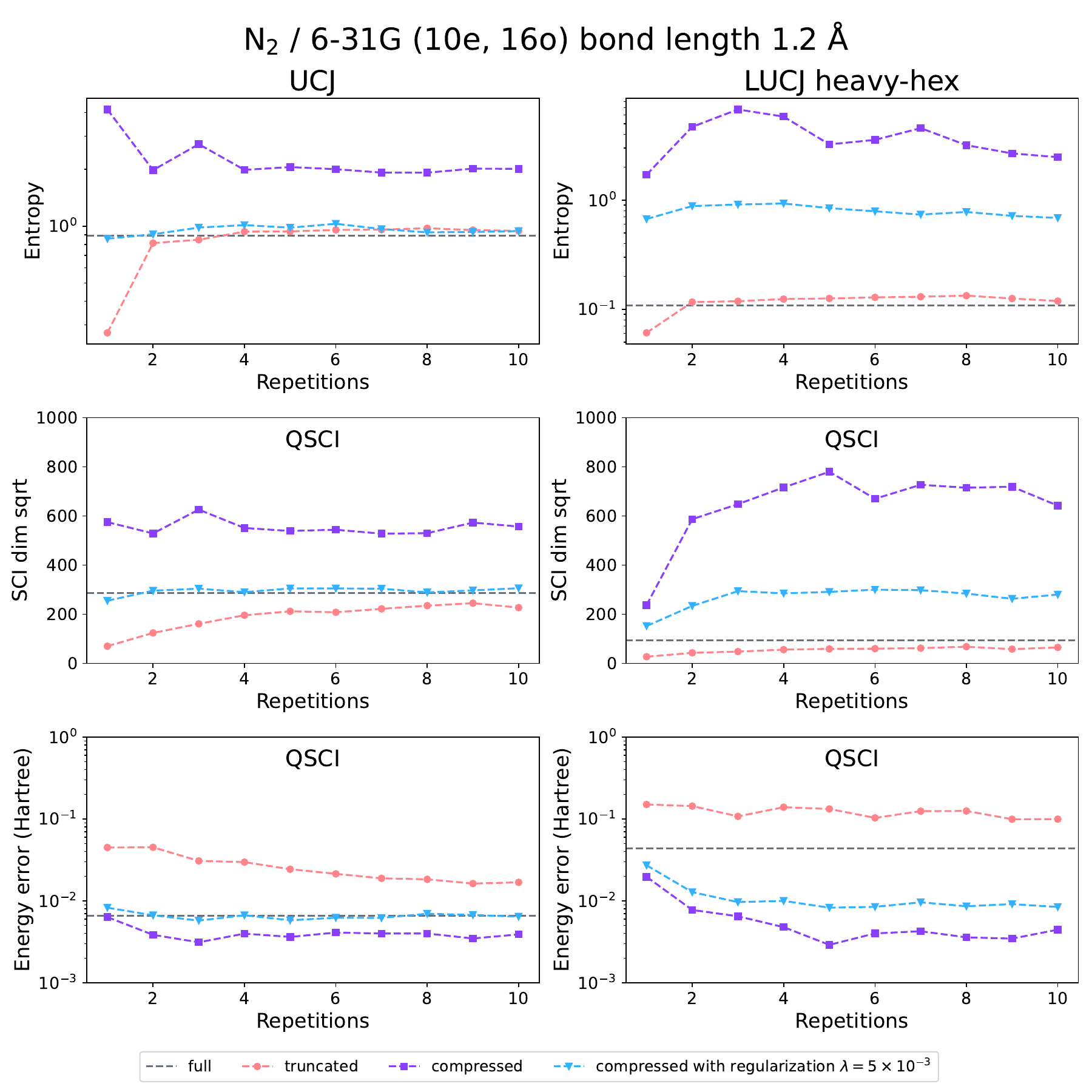}
    \caption{The compressed operator wavefunctions are less concentrated and yield more unique configurations when sampled. Data for the (10e,~16o) N\textsubscript{2} system at bond length 1.2 Å. The left column shows data for the UCJ ansatz, and the right column shows data for the LUCJ ansatz with heavy-hex connectivity. \textbf{Top row}: Entropy of the probability distribution associated with the state vector as a function of the number of ansatz repetitions, for naive truncation, compressed double-factorization, and compressed double-factorization with regularization. \textbf{Middle row}: Square root of the dimension of the subspace used for QSCI. The configurations of the subspace are formed by a Cartesian product, using the same single-spin configurations for both spin up and spin down. Generally, a higher-entropy distribution yields more unique configurations, and thus a larger subspace. The horizontal line indicates the entropy for the ansatz with the full untruncated number of repetitions. \textbf{Bottom row}: QSCI energy error. Generally, a higher subspace dimension yields better energy estimates. Compressed double-factorization can outperform even the full untruncated ansatz because the state vector can have higher entropy and produce more unique configurations when sampled.}
    \label{fig:entropy}
\end{figure}

Fig.~\ref{fig:regularization} shows the effect of adding the regularization term with 
$\lambda = 0.005$. 
Adding the regularization term substantially increases the loss function compared to without regularization, but it results in an improvement in the Hamiltonian expectation value over the naive truncation in almost all cases. However, regularization negatively impacts the QSCI energy. While increased operator norm degrades the Hamiltonian expectation value due to Trotter error, it may actually benefit QSCI by causing the wavefunction coefficients to spread out among a larger number of configurations, producing a more diverse set of samples. Fig.~\ref{fig:entropy} shows that the probability distribution associated with the wavefunction without regularization has a higher entropy than with regularization, leading to a higher subspace dimension used in QSCI. This example highlights the fact that expectation-value- and sample-based algorithms prefer different ansatz parameters.

Fig.~\ref{fig:regularization} also shows that all methods of parameter initialization yield larger errors at bond length 2.4 Å compared to 1.2 Å. As shown in the top row of Fig.~\ref{fig:energy_comp_and_truncation}, CCSD gives a non-variational energy value at 2.4 Å, signaling a breakdown of the CCSD approximation. Because the compressed double factorization targets the $t_2$ amplitudes from CCSD, it is inherently limited by the accuracy of CCSD itself.

\subsection{Parameter optimization for sample-based algorithms by tensor network simulation}
\label{sec:method:subsec:tn_opt}

In this section, we describe a method to improve the parameters of the LUCJ ansatz for sample-based algorithms using tensor network simulation. The idea is simple: use a tensor network simulator to approximately sample from the ansatz state, use those samples to run QSCI, and feed the resulting energy into a classical optimizer that tries to find parameters that minimize the energy. While the analogous idea has already been explored in the context of expectation value-based algorithms~\cite{rudolph2023synergistic,khan2023preoptimizing}, to our knowledge it has not yet been applied to sample-based algorithms like QSCI.

In this work, we begin the tensor network optimization with ansatz parameters obtained from the compressed double factorization of the CCSD $t$ amplitudes.
This method is applicable even in the absence of a physically motivated parameter initialization.
Starting with high-quality parameters can reduce the number of iterations needed in this classical surrogate optimization.
For the tensor network simulation we use matrix product states (MPS)~\cite{cirac2021mps} because they offer an effective balance between accuracy and computational cost. 
We also evaluated projected entangled pair states (PEPS)~\cite{cirac2021mps}, but
their increased representational power did not translate into improved accuracy for our workloads, while incurring significantly higher simulation cost. The higher cost may be justified for systems with different or more complex entanglement structure, and the use of alternative tensor network ansatzes and algorithms for simulating them is an active area of research~\cite{rudolph2026simulating}. We leave to future work the further exploration of this trade-off between tensor network simulation cost and accuracy.

Because the objective function involves converting a quantum circuit to an MPS, sampling from it, and running QSCI, the gradient is not readily available. Therefore, we choose to use derivative-free black box optimization, and in this work we use NOMAD~\cite{audet2022nomad}.
In general, black-box optimization methods rely on repeatedly querying the objective function to construct and solve an approximate model. 
As a result, they are typically most effective for problems with a moderate number of parameters, e.g., on the order of tens of parameters.
For our case with potentially hundreds of parameters, we found that NOMAD did not perform well, so we adopted PSD-MADS~\cite{audet2008psdmads} to decompose the optimization into subproblems that could each be effectively handled by NOMAD.

\subsection{Scaling considerations}

The compressed double factorization requires running CCSD to obtain the initial $t_2$-amplitudes, and the computational cost of CCSD scales as $O(N^6)$ where $N$ is the number of spatial orbitals~\cite{bartlett2007coupled}. Once the initial $t_2$-amplitudes are obtained, the compressed factorization runs a gradient-based optimizer for a capped number of iterations. Each iteration involves computing the least-squares objective~(\ref{eq:objective}) and its gradient. Since we use reverse-mode automatic differentiation, computing the gradient has the same asymptotic cost as the objective function itself~\cite{jax2018github}. Computing the objective function involves evaluating the $t_2$-amplitudes in Eq.~\ref{eq:double_factorization_t2}. Each entry of the amplitudes tensor requires summing $O(LN)$ values, and there are $O(N^4)$ entries, so the total cost of evaluating them is $O(LN^5)$. Thus, if the number of optimization iterations is capped at a constant, then the total cost of the compressed factorization is $O(LN^5)$. With $L$ typically being smaller than $N$, this scaling is no worse than that of the CCSD computation needed to obtain the initial $t_2$-amplitudes.

The parameter optimization by tensor network simulation involves running a gradient-free optimizer for a capped number of iterations, so its cost is governed by the cost of a single objective function evaluation. Evaluating the objective function involves simulating the LUCJ circuit with a tensor network, sampling configurations from the network, and diagonalizing the Hamiltonian in a subspace spanned by a set of sampled configurations, usually with an iterative eigensolver. Diagonalizing the Hamiltonian in a subspace of dimension $D$ costs $O(D N^4)$~\cite{kirby2026observation}. When the tensor network is an MPS, drawing $M$ samples by perfect sampling costs $O(MN\chi^2)$, where $\chi$ is the maximum bond dimension of the MPS~\cite{ferris2012perfect}. For systems with more complex entanglement structure, a more expressive tensor network ansatz such as PEPS may be preferable, but this comes at increased computational cost.

\section{Results}

This section presents simulation and hardware results for systems larger than the N\textsubscript{2}/6-31G system studied in Section~\ref{sec:methods}.

We perform noiseless simulations of N\textsubscript{2} in a (10e, 26o) active space derived from the cc-pVDZ basis set, as well as the iron-sulfur cluster [2Fe-2S] in a (30e, 20o) active space. At this scale, we can compute the exact wavefunction and sample configurations from it using \textsf{ffsim}~\cite{ffsim}, but we found computing the Hamiltonian expectation value to be prohibitively expensive (note that the N\textsubscript{2}/cc-pVDZ system would map to 52 qubits under standard fermion-to-qubit mappings~\cite{bravyi2002fermionic}). Therefore, we only present results for the QSCI energy. Due to the substantial cost of running FCI for these larger systems, we obtain our reference energy values instead using semistochastic heat-bath configuration interaction (SHCI)~\cite{sharma2017semistochastic,holmes2016heat}.

We perform hardware experiments for the N\textsubscript{2}/cc-pVDZ system, as well as a formamide dimer in a (36e, 30o) active space derived from the 6-31+G* basis set~\cite{cato2013exploring, frey2006ab}. The formamide system was beyond our capability for exact state vector simulation (storing the state vector alone would consume more than 100 petabytes of memory). Our experiments are conducted on a superconducting quantum processing unit (QPU) developed by IBM.

\subsection{Computational details}
\label{sec:computational_details}

\paragraph{Quantum chemistry}
We used \textsf{PySCF}~\cite{sun2018pyscf2,sun2020pyscf} to generate the molecular integrals for all systems except for [2Fe-2S], and to perform the FCI, CCSD, and CISD calculations for all systems. For the [2Fe-2S] system, we obtained the molecular integrals from Refs.~\cite{li2017github,li2017spinprojected}, who derived the active space from the TZP-DKH basis. We used Dice~\cite{sharma2017semistochastic,holmes2016heat} to run the SHCI calculations, with a value of $10^{-5}$ for the selection cutoff (the parameter referred to as $\epsilon_1$ in the Dice documentation~\cite{dicedocs}).

\paragraph{Compressed double factorization}
To perform the optimization for the compressed double factorization, we used the implementation of L-BFGS-B~\cite{byrd1995lbfgsb} in \textsf{SciPy}~\cite{2020SciPy-NMeth}, and set the maximum number of iterations to 100. We used \textsf{JAX}~\cite{jax2018github} to compute the objective function gradient with automatic differentiation. For the multi-stage optimization, we truncated by 2 repetitions in each iteration, starting at 20 repetitions for N\textsubscript{2}/6-31G and [2Fe-2S], and 50 for N\textsubscript{2}/cc-pVDZ.

\paragraph{State vector simulation and QSCI}
We used \textsf{ffsim}~\cite{ffsim} to compute the wavefunctions for the ansatz states, calculate the Hamiltonian expectation value of those states, and generate samples from them. We used \textsf{qiskit-addon-sqd}~\cite{qiskit-addon-sqd} to compute the QSCI energy from the samples. For each ansatz state, we sampled 100,000 configurations. Then, we subsampled 10 batches of 4,000 bitstrings uniformly at random. Each batch was used for an independent QSCI calculation, and we report the average, minimum, and maximum values from the batches. Note that the QSCI subspace is formed using a procedure that ensures it is symmetric with respect to swapping the labels of spin-up and spin-down orbitals~\cite{robledo2025sqd,qiskit-addon-sqd}, which is appropriate because all the systems we study have a closed-shell reference state.

\paragraph{Tensor network simulation}
We used \textsf{quimb}~\cite{gray2018quimb} for the MPS simulations with a maximum bond dimension 50, and a singular value cutoff of $1\cdot 10^{-10}$ for state truncation.
During tensor network optimization, we draw 10,000 samples from the MPS to perform QSCI.
To solve the black box optimization, 
we adopt the PSD-MADS algorithm implemented in \textsf{NOMAD}~\cite{audet2022nomad}
with a limit of 500 objective function evaluations.
The number of variables in the subproblem is set to 20 to be solved efficiently with the MADS algorithm. 
Four threads are available for solving the subproblems in parallel. 
For each subproblem, we set the limit of function evaluations to 20.

\paragraph{Hardware experiments and SQD}
We used \texttt{ibm\_kingston}, IBM's 156-qubit Heron r2 QPU, for the hardware experiments. We used the LUCJ ansatz tailored for the heavy-hex connectivity of the device, and we used one repetition of the ansatz. For each experiment, we performed 10 independent runs, each collecting 1,000,000 shots per circuit.
We selected a qubit layout based on a scoring heuristic which is the summation over the two-qubit gate error for all couplings and the measurement errors for the physical qubits in the pattern. The left panel of Figure~\ref{fig:hardware result} depicts the molecules studied, along with the corresponding physical qubit layouts. To suppress decoherence, we applied dynamical decoupling with the ``XY4'' pulse sequence. We used \textsf{qiskit-addon-sqd} to run QSCI and SQD, using the same settings as for the numerical simulations. For the results that include configuration recovery, we ran 10 iterations of configuration recovery and set the \texttt{carryover\_threshold} argument to $10^{-4}$.

\subsection{Compressed double factorization}
\label{sec:results:subsec:noiseless_sim}

\begin{figure}
    \centering
    \includegraphics[width=\linewidth]{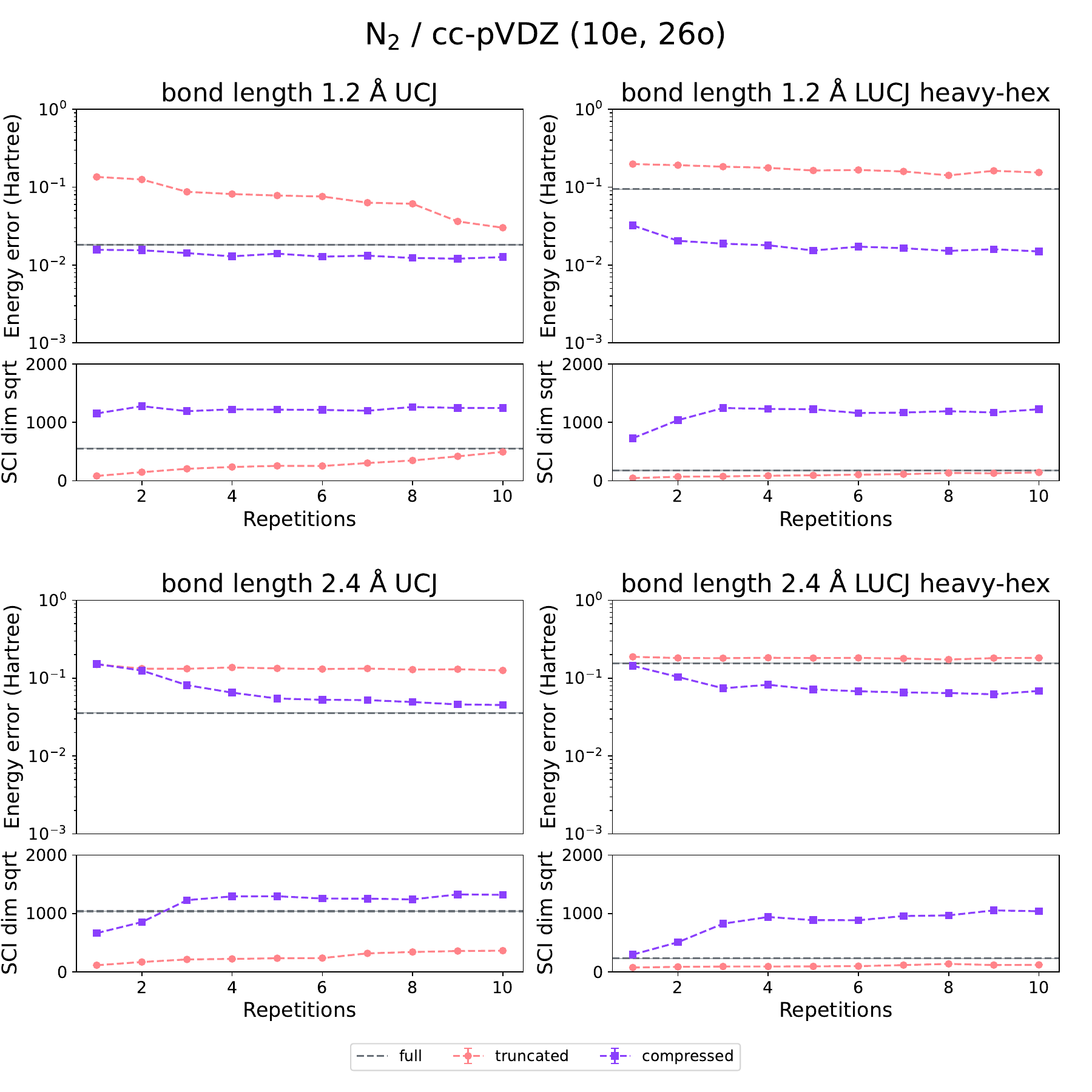}
    \caption{Compressed double factorization simulation results for N\textsubscript{2} in a (10e,~26o) active space derived from the cc-pVDZ basis set. Data for UCJ (left column) and LUCJ heavy-hex (right column), and bond lengths 1.2 Å (top row) and 2.4 Å (bottom row). The top panel of each plot shows the QSCI energy error, and the bottom panel shows the square root of the dimension of the subspace used for QSCI. For each data point, we sampled 100,000 bitstrings from the state vector, and then subsampled 10 batches of 4,000 bitstrings uniformly at random from which to construct the QSCI subspace. The data point shows the average value from the 10 batches, with error bars indicating the minimum and maximum values (when not visible, error bars are smaller than symbol sizes).}
    \label{fig:n2_cc-pvdz_sqd}
\end{figure}

\begin{figure}
    \centering
    \includegraphics[width=\linewidth]{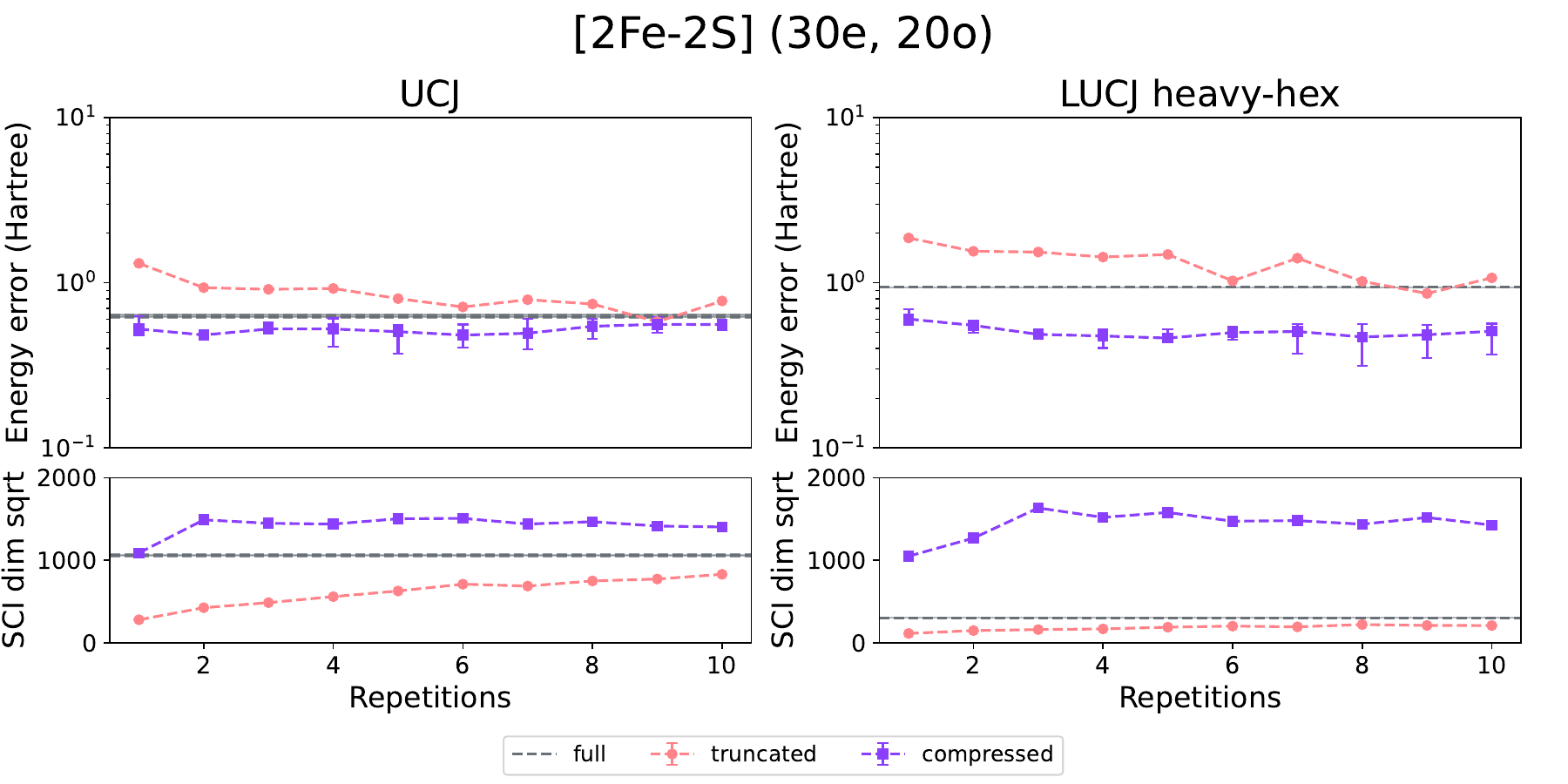}
    \caption{Compressed double factorization simulation results for 
    [2Fe-2S]
    in a (30e,~20o) active space. Data for UCJ (left column) and LUCJ heavy-hex (right column). The top panel of each plot shows the QSCI energy error, and the bottom panel shows the square root of the dimension of the subspace used for QSCI. For each data point, we sampled 100,000 bitstrings from the state vector, and then subsampled 10 batches of 4,000 bitstrings uniformly at random from which to construct the QSCI subspace. The data point shows the average value from the 10 batches, with error bars indicating the minimum and maximum values.}
    \label{fig:fe2s2_sqd}
\end{figure}

Fig.~\ref{fig:n2_cc-pvdz_sqd} presents results for the QSCI energy for the N\textsubscript{2}/cc-pVDZ (10e, 26o) system.
The results are consistent with our observations of the N\textsubscript{2}/6-31G system in Section~\ref{sec:method:subsec:compressed_double_factorization}.
The compressed operators deliver a significant improvement over the truncated operators in all cases,
and even outperform the LUCJ circuits with full layers in most cases.
With sparse connectivity, compressed operators do not exhibit performance degradation, 
highlighting the effectiveness of our heuristic in handling the connectivity constraint.
For bond length 1.2 Å, compressed operators yield a greater improvement over naive truncation compared with the results for bond length 2.4 Å. We attribute this to the fact that CCSD itself is more accurate at bond length 1.2 Å, which is near equilibrium.

Fig.~\ref{fig:fe2s2_sqd} presents results for the QSCI energy for the [2Fe-2S] (30e, 20o) system. Since CCSD did not converge when we attempted to run it, we use CISD instead to derive the $t_2$ amplitudes. Again, the parameters from the compressed double factorization consistently outperform the naive truncation. 
However, the absolute errors remain higher (greater than 0.1 Hartrees), and we attribute this to the suboptimality of CISD parameters.

\subsection{Parameter optimization by tensor network simulation}
\label{sec:results:subsec:tn_opt}

\begin{figure}
    \centering
    \includegraphics[width=0.6\linewidth]{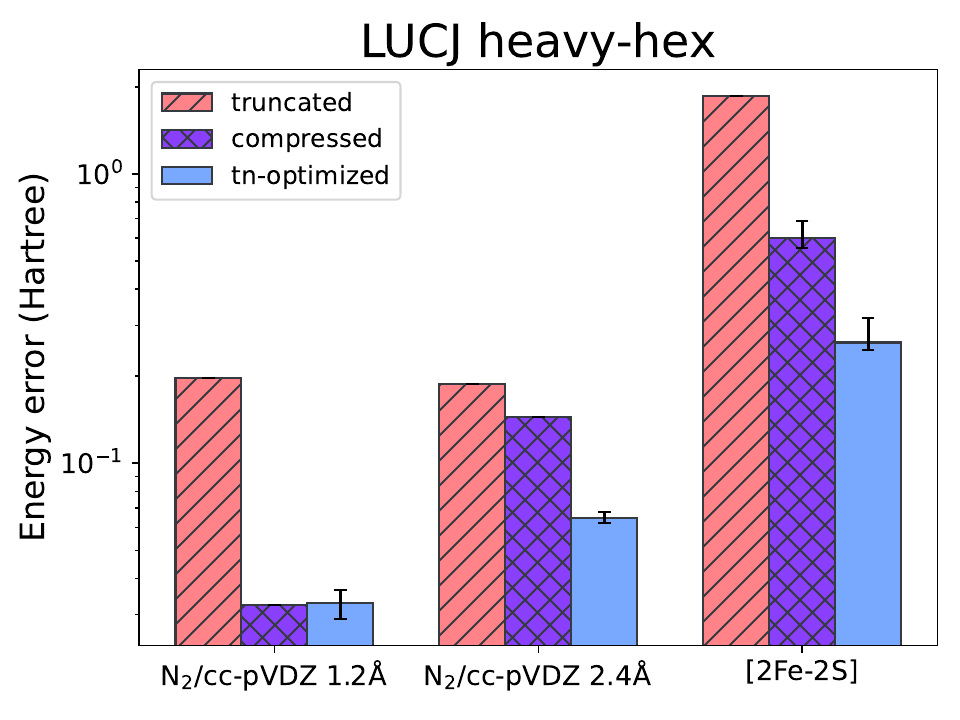}
    \caption{
    Comparison of the QSCI energy errors from naive truncation, compressed double-factorization, and tensor network optimization.
    Data is shown for N\textsubscript{2}/cc-pVDZ (10e,~26o) at two different bond lengths, and [2Fe-2S] (30e,~20o).
    The LUCJ ansatz was used with heavy-hex connectivity and a single repetition.
    For each data point, we sampled 100,000 bitstrings from the state vector, and then subsampled 10 batches of 4,000 bitstrings uniformly at random from which to construct the QSCI subspace. 
    The bar shows the average value from the 10 batches, with error bars indicating the minimum and maximum values.
    }
    \label{fig:tn_results}
\end{figure}

Fig.~\ref{fig:tn_results} shows the QSCI energy of the parameters obtained after tensor network optimization, and compares it to the energies obtained with naive truncation and compressed double factorization. While we do not see a significant improvement for N\textsubscript{2}/cc-pVDZ at bond length 1.2 Å, we do see significant improvements for N\textsubscript{2}/cc-pVDZ at 2.4 Å and [2Fe-2S]. We attribute the greater improvements in these latter systems to the worse performance of CCSD leaving more room for improvement. Indeed, as mentioned previously, we initialized the [2Fe-2S] system from CISD because CCSD did not converge.

\subsection{Hardware results}
\label{sec:results:subsec:hardware_sim}

For our hardware experiments, we compared four different ways of initializing the LUCJ parameters:
\begin{itemize}
    \item Random. Here, orbital rotations are initialized from Haar-random unitaries, and the diagonal Coulomb matrices have entries sampled uniformly from the range $[-\pi, \pi]$. We include this case as a control, to ensure that we are actually able to extract a signal from the noisy QPUs.
    \item Truncated. This is the naive truncation strategy.
    \item Compressed. This is the compressed double factorization.
    \item Zeros (no-LUCJ). Here, the Hartree-Fock state is directly prepared and sampled, without applying an LUCJ circuit. This is equivalent to setting all the gate angles in the LUCJ ansatz to zero.
\end{itemize}

\begin{figure}
    \centering
    \includegraphics[width=1.0\linewidth]{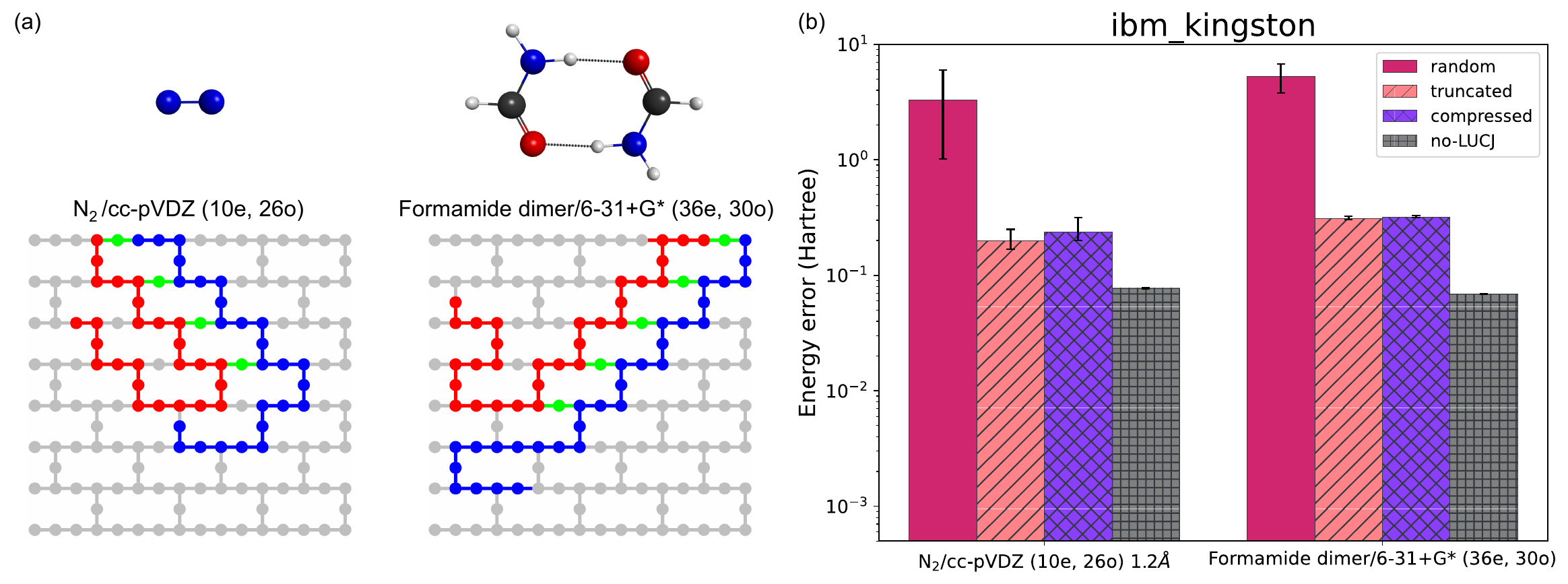}
    \caption{(a) Qubit layouts of LUCJ circuits on \texttt{ibm\_kingston} for each molecule. Device qubits are shown in gray; qubits encoding spin-up/down electron occupations are red/blue, and ancilla qubits are green. Molecular structures above correspond to the qubit layouts, with C, O, N, H atoms in black, red, blue, and gray, respectively. (b) QSCI energy error (in Hartree) obtained using the LUCJ ansatz with different parameter initialization schemes—random, truncated, compressed, and no-LUCJ—for both molecules. Error bars represent the range of energy errors across ten independent runs.}
    \label{fig:hardware result}
\end{figure}

The right panel of Figure~\ref{fig:hardware result} shows the QSCI energy error for the four different settings. In both molecular systems, the truncated and compressed parameter settings achieve an order of magnitude lower error compared to random parameters. There is not a significant difference between the truncated and compressed parameters, an observation that we attribute to hardware noise. The no-LUCJ initialization gives the lowest QSCI energy error. We attribute this behavior to the shallow state-preparation circuit and to hardware errors generating additional configurations near Hartree-Fock, as explained below.

In the occupation encoding used here, the Hartree-Fock state is a single bitstring in the computational basis. In the absence of noise, sampling this state would yield only the Hartree-Fock determinant, and projecting and diagonalizing the Hamiltonian in that subspace of dimension one would return the Hartree-Fock energy. The no-LUCJ circuit has no two-qubit gates, but state-preparation and measurement errors can still change the observed occupation bits. After postselecting configurations with the correct electron numbers in both spin sectors, some of these additional configurations can be useful excited determinants. QSCI can then recover correlation energy by diagonalizing the Hamiltonian in the subspace constructed from the samples. Thus, the no-LUCJ QSCI energy can differ from the Hartree-Fock energy even though no correlation-generating LUCJ circuit is applied. The expectation that these additional configurations tend to be low-rank excitations can be understood qualitatively using a weak independent bit-flip model. A particle-number-preserving single excitation requires changing one occupied bit to zero and one unoccupied bit to one, hence two flips. Double and triple excitations require four and six flips, respectively. If each occupation bit flips independently with probability $p\ll 1$, the probability of a specified determinant at excitation rank $k$ is of order $p^{2k}$. This suppresses individual higher-rank excitations, although their total probability also depends on the number of possible configurations at each rank. This is a qualitative explanation, and correlated errors can modify this picture.

\begin{figure}
    \centering
    \includegraphics[width=1\linewidth]{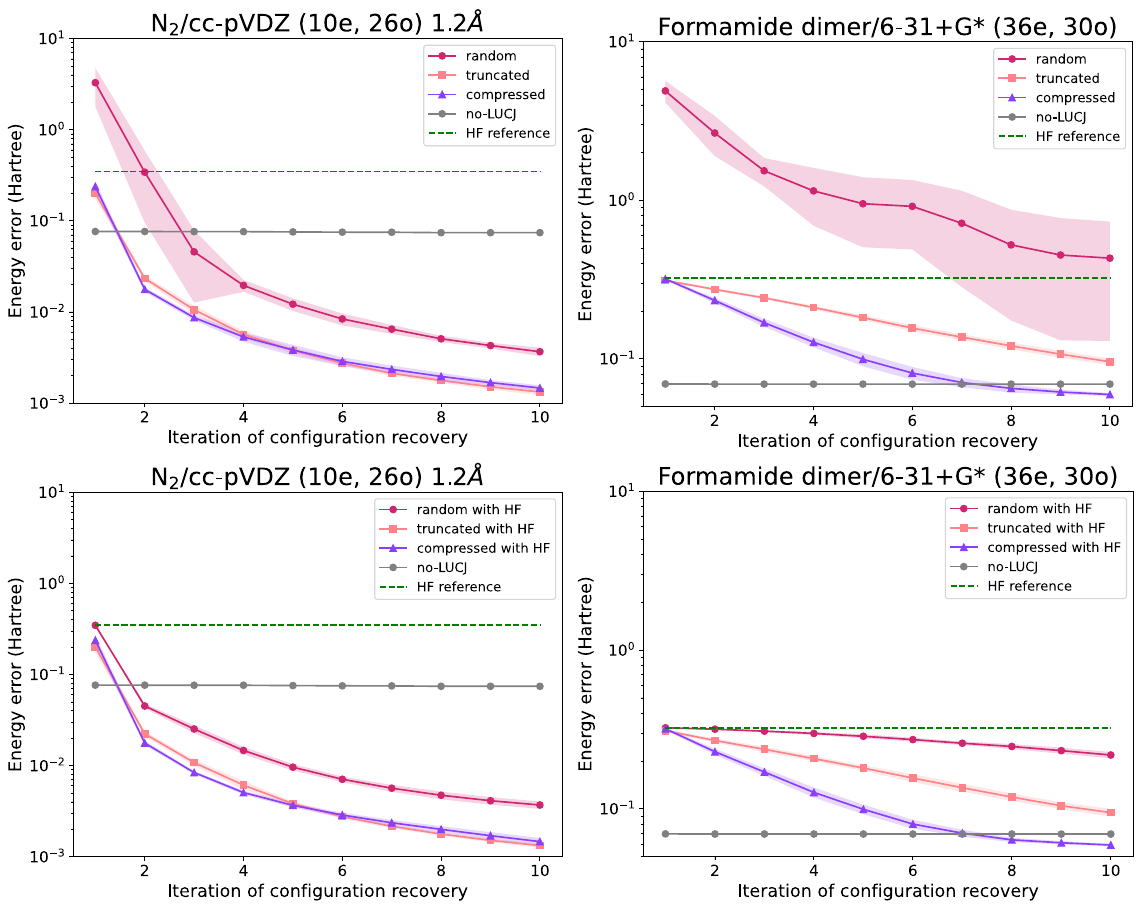}
    \caption{Convergence of SQD energy error during configuration recovery. The plots show the energy error (Hartree, logarithmic scale) as a function of the iteration of configuration recovery for two molecular systems: N\textsubscript{2}/cc-pVDZ (10e, 26o) at bond length 1.2 Å and formamide dimer/6-31+G* (36e, 30o). Four sampling strategies are compared: random, truncated, compressed, and no-LUCJ. The solid lines represent the mean error across multiple runs, and the shaded regions indicate the standard deviation. The ``no-LUCJ'' line (gray) corresponds to sampling the Hartree-Fock state directly on the QPU without applying a LUCJ circuit. The ``HF reference'' (green dashed line) marks the classical Hartree-Fock energy.
    \textbf{Top row}: Results of standard SQD.
    \textbf{Bottom row}: Results of SQD where the Hartree-Fock determinant is explicitly included in the configuration subspace during diagonalization.}
    \label{fig:configuration-recovery}
\end{figure}

We ran configuration recovery to obtain SQD energies, and the results are shown in Figure~\ref{fig:configuration-recovery}. Because of the greatly reduced hardware noise in the no-LUCJ case, configuration recovery does not lower the error further. Configuration recovery does lower the error of the other initialization methods, and for the N\textsubscript{2} system, they all achieve lower error than no-LUCJ. For the formamide dimer, the compressed initialization ends up achieving lower error than the truncated initialization, and is the only method that achieves lower error than no-LUCJ within 10 iterations. For the N\textsubscript{2} system, we still do not observe an advantage of the compressed initialization over the truncated. We attribute this to the N\textsubscript{2}/cc-pVDZ system being more challenging for noisy quantum hardware due to its highly unbalanced electron-to-orbital ratio making it more difficult to sample valid bitstrings that have the correct number of electrons in both spin sectors. As shown in Table~\ref{tab:sample_statistic-2Q}, in the N\textsubscript{2} system, we sample an order of magnitude fewer valid bitstrings compared to the formamide dimer. Nevertheless, our results show that the compressed parameter initialization can outperform the naive truncated parameters on real quantum hardware, and that either is very likely to outperform random parameters.

Beyond configuration recovery, quantum error mitigation can also be extended to bitstring sampling. Ref.~\cite{liu2025samplingerrormitigation} shows how techniques such as probabilistic error cancellation and zero-noise extrapolation~\cite{temme2017errormitigation} can estimate computational-basis probabilities from signed measurement statistics obtained using additional circuit variants. They also describe how to sample from the resulting corrected distribution by adding and removing samples according to these signs. The sampling overhead depends on the mitigation technique and noise level, and finite-shot estimates can contain negative probabilities that require suitable treatment before resampling. Applying such techniques to QSCI or SQD could alter the probabilities used to select configurations, but a more accurate sampling distribution does not by itself guarantee a lower energy, since the energy is obtained by diagonalizing the Hamiltonian in the selected subspace. While we did not test these techniques here, assessing their cost and benefit for QSCI and SQD is an interesting direction for future work.

\begin{table}
\resizebox{\columnwidth}{!}{%
\begin{tabular}{c|c|c|c|c|c}
\hline \hline
\multirow{2}{*}{system}& \multicolumn{1}{c|}{\multirow{2}{*}{param initialization}}&\multicolumn{2}{|c|}{quantum resources}          & \multicolumn{2}{l}{\#valid bitstrings} \\ 
\cline{3-6}& \multicolumn{1}{|c|}{}                         & \ \ \ \ 2Q depth\ \ \ \ \  & 2Q gate count & avg                       & std dev                          \\
\hline

\multirow{4}{*}{N\textsubscript{2}/cc-pVDZ (10e, 26o), 1.2 Å}
& random&114&1858& 752& 316\\
& truncated&110&1726& 1075& 434\\
& compressed&114& 1858& 806& 365\\
& no-LUCJ& 0 & 0 & 315 & 31\\\hline

\multirow{4}{*}{Formamide dimer (36e, 30o)}\
& random    & 156& 2838& 8430 & 493\\
& truncated & 126& 2684& 6832& 173\\
& compressed& 156& 2838& 6723& 2934
\\
& no-LUCJ& 0 & 0 & 1013 &158\\\hline
\hline
\end{tabular}
}
\caption{Details of hardware experiments. For each system and parameter initialization method, we list the two-qubit gate depth and gate counts for the transpiled circuit executed on the QPU, as well as the average number (and standard deviation) of unique valid bitstrings (that is, configurations with the correct number of particles in both spin sectors) that were sampled.}
\label{tab:sample_statistic-2Q}
\end{table}

\section{Conclusion}
\label{sec:conclusion}

In this work, we introduced two complementary methods to improve parameter initialization for the LUCJ ansatz, a quantum circuit ansatz for chemistry applications. The first method employs compressed double factorization to better approximate the $t_2$ amplitudes from a CCSD calculation, and the second method uses approximate tensor network simulation to optimize the parameters of the ansatz for sample-based algorithms like QSCI. We evaluated our methods using both exact classical state vector simulation and experiments on superconducting quantum processors.

Our simulation results demonstrate that compressed double factorization significantly improves upon naive truncation across multiple molecular systems. In the case of the QSCI energy, it can even outperform the untruncated ansatz because the ansatz state becomes less concentrated (higher entropy) and produces a more diverse set of configurations when sampled. The tensor network optimization provided additional energy improvements beyond compressed double factorization, particularly for systems where the initial CCSD or CISD approximation is less accurate. Our hardware results show that the improved parameters for the LUCJ ansatz can make a difference on the noisy QPUs that are available today.

Together, these methods developed in this work offer a practical way to improve the performance of variational quantum algorithms that use the UCJ or LUCJ ansatzes. By carefully balancing the trade-offs between physical accuracy, hardware constraints, and the specific requirements of expectation value- and sample-based algorithms, our approaches enable more accurate quantum chemistry calculations on near-term quantum processors.


\section{Author contributions}

KJS conceived the project. WL and KJS wrote the code implementation and performed the numerical simulations, with assistance from MM and HZ. FL collected and analyzed the QPU data. All authors contributed to writing the manuscript, and to discussions that shaped the project. Generative AI was used to write some code and parts of the manuscript.

\section{Acknowledgments}
\label{sec:acknowledgment}

We thank Abdullah Ash Saki for providing code to pick good qubit layouts for the LUCJ circuits on the IBM QPUs. WL is grateful for the mentorship of Nate Earnest-Noble during her summer internship at IBM Quantum. FL acknowledges Thaddeus Pellegrini from the Quantum Algorithm Engineering team at IBM for valuable discussions and coordinating with the IBM hardware team to optimize circuit execution on the IBM QPUs. Some parts of this paper are adapted from parts of the \textsf{ffsim} documentation~\cite{ffsim} written by KJS and WL.

\section{Code and data availability}

The source code for this work is available at \url{https://github.com/kevinsung/lucj}. The hardware data (bitstrings collected from the QPUs) is available at \url{https://zenodo.org/records/17704385}. We have contributed an implementation of the compressed double factorization to the open-source software library \textsf{ffsim}~\cite{ffsim}.

\bibliographystyle{quantum}
\bibliography{ref}

\begin{thebibliography}{10}

\bibitem{mcardle2020quantum}
Sam McArdle, Suguru Endo, Al\'an Aspuru-Guzik, Simon~C. Benjamin, and Xiao
  Yuan.
\newblock ``Quantum computational chemistry''.
\newblock \href{https://dx.doi.org/10.1103/RevModPhys.92.015003}{Rev. Mod.
  Phys. {\bf 92}, 015003}~(2020).

\bibitem{motta2022emerging}
Mario Motta and Julia~E. Rice.
\newblock ``Emerging quantum computing algorithms for quantum chemistry''.
\newblock \href{https://dx.doi.org/10.1002/wcms.1580}{WIREs Computational
  Molecular Science {\bf 12}, e1580}~(2022).

\bibitem{yoshioka2025krylov}
Nobuyuki Yoshioka, Mirko Amico, William Kirby, Petar Jurcevic, Arkopal Dutt,
  Bryce Fuller, Shelly Garion, Holger Haas, Ikko Hamamura, Alexander Ivrii,
  et~al.
\newblock ``Krylov diagonalization of large many-body hamiltonians on a quantum
  processor''.
\newblock \href{https://dx.doi.org/10.1038/s41467-025-59716-z}{Nature
  Communications {\bf 16}, 5014}~(2025).

\bibitem{yu2025quantumcentricalgorithmsamplebasedkrylov}
Jeffery Yu, Javier~Robledo Moreno, Joseph~T. Iosue, Luke Bertels, Daniel
  Claudino, Bryce Fuller, Peter Groszkowski, Travis~S. Humble, Petar Jurcevic,
  William Kirby, Thomas~A. Maier, Mario Motta, Bibek Pokharel, Alireza Seif,
  Amir Shehata, Kevin~J. Sung, Minh~C. Tran, Vinay Tripathi, Antonio Mezzacapo,
  and Kunal Sharma.
\newblock ``Quantum-centric algorithm for sample-based {Krylov}
  diagonalization''~(2025).
\newblock  \href{http://arxiv.org/abs/2501.09702}{arXiv:2501.09702}.

\bibitem{peruzzo2014variational}
Alberto Peruzzo, Jarrod McClean, Peter Shadbolt, Man-Hong Yung, Xiao-Qi Zhou,
  Peter~J. Love, Al{\'a}n Aspuru-Guzik, and Jeremy~L. O'Brien.
\newblock ``A variational eigenvalue solver on a photonic quantum processor''.
\newblock \href{https://dx.doi.org/10.1038/ncomms5213}{Nature Communications
  {\bf 5}, 4213}~(2014).

\bibitem{kanno2026qsci}
Keita Kanno, Masaya Kohda, Ryosuke Imai, Sho Koh, Kosuke Mitarai, Wataru
  Mizukami, and Yuya~O. Nakagawa.
\newblock ``Quantum-selected configuration interaction: Classical
  diagonalization of {Hamiltonians} in subspaces selected by quantum
  computers''.
\newblock \href{https://dx.doi.org/10.1103/dmn4-snfx}{Phys. Rev. Res. {\bf 8},
  023268}~(2026).

\bibitem{anand2022uccsd}
Abhinav Anand, Philipp Schleich, Sumner Alperin-Lea, Phillip~WK Jensen, Sukin
  Sim, Manuel D{\'\i}az-Tinoco, Jakob~S Kottmann, Matthias Degroote, Artur~F
  Izmaylov, and Al{\'a}n Aspuru-Guzik.
\newblock ``A quantum computing view on unitary coupled cluster theory''.
\newblock \href{https://dx.doi.org/10.1039/D1CS00932J}{Chemical Society Reviews
  {\bf 51}, 1659--1684}~(2022).

\bibitem{motta2020quantum_simulation_of_electronic_structure_with_transcorrelated_hamiltonian}
Mario Motta, Tanvi~P. Gujarati, Julia~E. Rice, Ashutosh Kumar, Conner Masteran,
  Joseph~A. Latone, Eunseok Lee, Edward~F. Valeev, and Tyler~Y. Takeshita.
\newblock ``Quantum simulation of electronic structure with a transcorrelated
  {Hamiltonian}: improved accuracy with a smaller footprint on the quantum
  computer''.
\newblock \href{https://dx.doi.org/10.1039/D0CP04106H}{Phys. Chem. Chem. Phys.
  {\bf 22}, 24270--24281}~(2020).

\bibitem{kandala2017hardwareefficient}
Abhinav Kandala, Antonio Mezzacapo, Kristan Temme, Maika Takita, Markus Brink,
  Jerry~M. Chow, and Jay~M. Gambetta.
\newblock ``Hardware-efficient variational quantum eigensolver for small
  molecules and quantum magnets''.
\newblock \href{https://dx.doi.org/10.1038/nature23879}{Nature {\bf 549},
  242--246}~(2017).

\bibitem{larocca2025barren}
Mart{\'i}n Larocca, Supanut Thanasilp, Samson Wang, Kunal Sharma, Jacob
  Biamonte, Patrick~J. Coles, Lukasz Cincio, Jarrod~R. McClean, Zo{\"e} Holmes,
  and M.~Cerezo.
\newblock ``Barren plateaus in variational quantum computing''.
\newblock \href{https://dx.doi.org/10.1038/s42254-025-00813-9}{Nature Reviews
  Physics {\bf 7}, 174--189}~(2025).

\bibitem{Dcunha2023hardware_efficient_ansatze}
Ruhee D’Cunha, T.~Daniel Crawford, Mario Motta, and Julia~E. Rice.
\newblock ``Challenges in the use of quantum computing hardware-efficient
  ansätze in electronic structure theory''.
\newblock \href{https://dx.doi.org/10.1021/acs.jpca.2c08430}{The Journal of
  Physical Chemistry A {\bf 127}, 3437--3448}~(2023).

\bibitem{matsuzawa2020ucj}
Yuta Matsuzawa and Yuki Kurashige.
\newblock ``Jastrow-type decomposition in quantum chemistry for low-depth
  quantum circuits''.
\newblock \href{https://dx.doi.org/10.1021/acs.jctc.9b00963}{Journal of
  Chemical Theory and Computation {\bf 16}, 944--952}~(2020).

\bibitem{motta2023lucj}
Mario Motta, Kevin~J. Sung, K.~Birgitta Whaley, Martin Head-Gordon, and James
  Shee.
\newblock ``Bridging physical intuition and hardware efficiency for correlated
  electronic states: the local unitary cluster jastrow ansatz for electronic
  structure''.
\newblock \href{https://dx.doi.org/10.1039/D3SC02516K}{Chem. Sci. {\bf 14},
  11213--11227}~(2023).

\bibitem{blunt2025quantum}
Nick~S. Blunt, Laura Caune, and Javiera Quiroz-Fernandez.
\newblock ``Quantum computing approach to fixed-node {Monte Carlo} using
  classical shadows''.
\newblock \href{https://dx.doi.org/10.1021/acs.jctc.4c01468}{Journal of
  Chemical Theory and Computation {\bf 21}, 1652--1666}~(2025).

\bibitem{liepuoniute2025quantumcentric}
Ieva Liepuoniute, Kirstin~D. Doney, Javier Robledo~Moreno, Joshua~A. Job,
  William~S. Friend, and Gavin~O. Jones.
\newblock ``Quantum-centric computational study of methylene singlet and
  triplet states''.
\newblock \href{https://dx.doi.org/10.1021/acs.jctc.5c00075}{Journal of
  Chemical Theory and Computation {\bf 21}, 5062--5070}~(2025).

\bibitem{kaliakin2025implicit_solvent}
Danil Kaliakin, Akhil Shajan, Fangchun Liang, and Kenneth M.~Jr. Merz.
\newblock ``Implicit solvent sample-based quantum diagonalization''.
\newblock \href{https://dx.doi.org/10.1021/acs.jpcb.5c01030}{The Journal of
  Physical Chemistry B {\bf 129}, 5788--5796}~(2025).

\bibitem{robledo2025sqd}
Javier Robledo-Moreno, Mario Motta, Holger Haas, Ali Javadi-Abhari, Petar
  Jurcevic, William Kirby, Simon Martiel, Kunal Sharma, Sandeep Sharma,
  Tomonori Shirakawa, Iskandar Sitdikov, Rong-Yang Sun, Kevin~J. Sung, Maika
  Takita, Minh~C. Tran, Seiji Yunoki, and Antonio Mezzacapo.
\newblock ``Chemistry beyond the scale of exact diagonalization on a
  quantum-centric supercomputer''.
\newblock \href{https://dx.doi.org/10.1126/sciadv.adu9991}{Science Advances
  {\bf 11}, eadu9991}~(2025).

\bibitem{shajan2025toward}
Akhil Shajan, Danil Kaliakin, Abhishek Mitra, Javier Robledo~Moreno, Zhen Li,
  Mario Motta, Caleb Johnson, Abdullah~Ash Saki, Susanta Das, Iskandar
  Sitdikov, Antonio Mezzacapo, and Kenneth~M. Merz.
\newblock ``Toward quantum-centric simulations of extended molecules:
  Sample-based quantum diagonalization enhanced with density matrix embedding
  theory''.
\newblock \href{https://dx.doi.org/10.1021/acs.jctc.5c00114}{Journal of
  Chemical Theory and Computation {\bf 21}, 6801--6810}~(2025).

\bibitem{bartlett2007coupled}
Rodney~J. Bartlett and Monika Musia\l{}.
\newblock ``Coupled-cluster theory in quantum chemistry''.
\newblock \href{https://dx.doi.org/10.1103/RevModPhys.79.291}{Rev. Mod. Phys.
  {\bf 79}, 291--352}~(2007).

\bibitem{Cohn_2021_compressed_df}
Jeffrey Cohn, Mario Motta, and Robert~M. Parrish.
\newblock ``Quantum filter diagonalization with compressed double-factorized
  {Hamiltonians}''.
\newblock \href{https://dx.doi.org/10.1103/PRXQuantum.2.040352}{PRX Quantum
  {\bf 2}, 040352}~(2021).

\bibitem{2023ClassicalSurrogates}
Franz~J. Schreiber, Jens Eisert, and Johannes~Jakob Meyer.
\newblock ``Classical surrogates for quantum learning models''.
\newblock \href{https://dx.doi.org/10.1103/PhysRevLett.131.100803}{Phys. Rev.
  Lett. {\bf 131}, 100803}~(2023).

\bibitem{hirsbrunner2024beyond}
Mark~R. Hirsbrunner, Diana Chamaki, J.~Wayne Mullinax, and Norm~M. Tubman.
\newblock ``Beyond {MP}2 initialization for unitary coupled cluster quantum
  circuits''.
\newblock \href{https://dx.doi.org/10.22331/q-2024-11-26-1538}{{Quantum} {\bf
  8}, 1538}~(2024).

\bibitem{2025PNAS}
Erik~J. Gustafson, Juha Tiihonen, Diana Chamaki, Farshud Sorourifar, J.~Wayne
  Mullinax, Andy C.~Y. Li, Filip~B. Maciejewski, Nicolas P.~D. Sawaya, Jaron~T.
  Krogel, David E.~Bernal Neira, and Norm~M. Tubman.
\newblock ``Surrogate optimization of variational quantum circuits''.
\newblock \href{https://dx.doi.org/10.1073/pnas.2408530122}{Proceedings of the
  National Academy of Sciences {\bf 122}, e2408530122}~(2025).

\bibitem{rudolph2023synergistic}
Manuel~S Rudolph, Jacob Miller, Danial Motlagh, Jing Chen, Atithi Acharya, and
  Alejandro Perdomo-Ortiz.
\newblock ``Synergistic pretraining of parametrized quantum circuits via tensor
  networks''.
\newblock \href{https://dx.doi.org/10.1038/s41467-023-43908-6}{Nature
  Communications {\bf 14}, 8367}~(2023).

\bibitem{khan2023preoptimizing}
Abid Khan, Bryan~K. Clark, and Norm~M. Tubman.
\newblock ``Pre-optimizing variational quantum eigensolvers with tensor
  networks''~(2023).
\newblock  \href{http://arxiv.org/abs/2310.12965}{arXiv:2310.12965}.

\bibitem{liu2023mitigating}
Huan-Yu Liu, Tai-Ping Sun, Yu-Chun Wu, Yong-Jian Han, and Guo-Ping Guo.
\newblock ``Mitigating barren plateaus with transfer-learning-inspired
  parameter initializations''.
\newblock \href{https://dx.doi.org/10.1088/1367-2630/acb58e}{New Journal of
  Physics {\bf 25}, 013039}~(2023).

\bibitem{park2024hamiltonian}
Chae-Yeun Park and Nathan Killoran.
\newblock ``Hamiltonian variational ansatz without barren plateaus''.
\newblock \href{https://dx.doi.org/10.22331/q-2024-02-01-1239}{{Quantum} {\bf
  8}, 1239}~(2024).

\bibitem{mhiri2025unifyingaccountwarmstart}
Hela Mhiri, Ricard Puig, Sacha Lerch, Manuel~S. Rudolph, Thiparat Chotibut,
  Supanut Thanasilp, and Zoë Holmes.
\newblock ``A unifying account of warm start guarantees for patches of quantum
  landscapes''~(2025).
\newblock  \href{http://arxiv.org/abs/2502.07889}{arXiv:2502.07889}.

\bibitem{ffsim}
Kevin~J. Sung, Inho Choi, Mirko Amico, Bartholomew Andrews, Esra Ayantuna,
  Yukio Kawashima, Wan-Hsuan Lin, David Omanovic, Samuele Piccinelli,
  Javier~Robledo Moreno, Abdullah~Ash Saki, James Shee, Soyoung Shin, Minh~C.
  Tran, Kento Ueda, Haimeng Zhang, and Mario Motta.
\newblock ``ffsim: Faster simulation of fermionic quantum circuits''~(2026).
\newblock  \href{http://arxiv.org/abs/2605.03123}{arXiv:2605.03123}.

\bibitem{wecker2015progress}
Dave Wecker, Matthew~B. Hastings, and Matthias Troyer.
\newblock ``Progress towards practical quantum variational algorithms''.
\newblock \href{https://dx.doi.org/10.1103/PhysRevA.92.042303}{Phys. Rev. A
  {\bf 92}, 042303}~(2015).

\bibitem{harsha2018on_the_difference}
Gaurav Harsha, Toru Shiozaki, and Gustavo~E. Scuseria.
\newblock ``On the difference between variational and unitary coupled cluster
  theories''.
\newblock \href{https://dx.doi.org/10.1063/1.5011033}{The Journal of Chemical
  Physics {\bf 148}, 044107}~(2018).

\bibitem{raghavachari1989afifthorder}
Krishnan Raghavachari, Gary~W. Trucks, John~A. Pople, and Martin Head-Gordon.
\newblock ``A fifth-order perturbation comparison of electron correlation
  theories''.
\newblock \href{https://dx.doi.org/10.1016/S0009-2614(89)87395-6}{Chemical
  Physics Letters {\bf 157}, 479--483}~(1989).

\bibitem{motta2021lowrank}
Mario Motta, Erika Ye, Jarrod~R. McClean, Zhendong Li, Austin~J. Minnich, Ryan
  Babbush, and Garnet Kin-Lic Chan.
\newblock ``Low rank representations for quantum simulation of electronic
  structure''.
\newblock \href{https://dx.doi.org/10.1038/s41534-021-00416-z}{npj Quantum
  Information {\bf 7}, 83}~(2021).

\bibitem{moreno2023enhancing}
Javier~Robledo Moreno, Jeffrey Cohn, Dries Sels, and Mario Motta.
\newblock ``Enhancing the expressivity of variational neural, and
  hardware-efficient quantum states through orbital rotations''~(2023).
\newblock  \href{http://arxiv.org/abs/2302.11588}{arXiv:2302.11588}.

\bibitem{sherrill1999ci}
C.~{David Sherrill} and Henry~F. Schaefer.
\newblock ``The configuration interaction method: Advances in highly correlated
  approaches''.
\newblock \href{https://dx.doi.org/10.1016/S0065-3276(08)60532-8}{Advances in
  Quantum Chemistry {\bf 34}, 143--269}~(1999).

\bibitem{hino2006tailored}
Osamu Hino, Tomoko Kinoshita, Garnet Kin-Lic Chan, and Rodney~J. Bartlett.
\newblock ``Tailored coupled cluster singles and doubles method applied to
  calculations on molecular structure and harmonic vibrational frequencies of
  ozone''.
\newblock \href{https://dx.doi.org/10.1063/1.2180775}{The Journal of Chemical
  Physics {\bf 124}, 114311}~(2006).

\bibitem{morchen2020tailored}
Maximilian Mörchen, Leon Freitag, and Markus Reiher.
\newblock ``Tailored coupled cluster theory in varying correlation regimes''.
\newblock \href{https://dx.doi.org/10.1063/5.0032661}{The Journal of Chemical
  Physics {\bf 153}, 244113}~(2020).

\bibitem{scheurer2024tailored}
Maximilian Scheurer, Gian-Luca~R. Anselmetti, Oumarou Oumarou, Christian
  Gogolin, and Nicholas~C. Rubin.
\newblock ``Tailored and externally corrected coupled cluster with quantum
  inputs''.
\newblock \href{https://dx.doi.org/10.1021/acs.jctc.4c00037}{Journal of
  Chemical Theory and Computation {\bf 20}, 5068--5093}~(2024).

\bibitem{byrd1995lbfgsb}
Richard~H. Byrd, Peihuang Lu, Jorge Nocedal, and Ciyou Zhu.
\newblock ``A limited memory algorithm for bound constrained optimization''.
\newblock \href{https://dx.doi.org/10.1137/0916069}{SIAM J. Sci. Comp {\bf 16},
  1190--1208}~(1995).

\bibitem{cirac2021mps}
J.~Ignacio Cirac, David P\'erez-Garc\'{\i}a, Norbert Schuch, and Frank
  Verstraete.
\newblock ``Matrix product states and projected entangled pair states:
  Concepts, symmetries, theorems''.
\newblock \href{https://dx.doi.org/10.1103/RevModPhys.93.045003}{Rev. Mod.
  Phys. {\bf 93}, 045003}~(2021).

\bibitem{rudolph2026simulating}
Manuel~S. Rudolph and Joseph Tindall.
\newblock ``Simulating and sampling from quantum circuits with {2D} tensor
  networks''.
\newblock \href{https://dx.doi.org/10.1038/s41534-026-01372-2}{npj Quantum
  Information}~(2026).

\bibitem{audet2022nomad}
Charles Audet, S\'{e}bastien Le~Digabel, Viviane~Rochon Montplaisir, and
  Christophe Tribes.
\newblock ``{Algorithm 1027: NOMAD Version 4: Nonlinear Optimization with the
  MADS Algorithm}''.
\newblock \href{https://dx.doi.org/10.1145/3544489}{ACM Trans. Math. Softw.
  {\bf 48}, 35}~(2022).

\bibitem{audet2008psdmads}
Charles Audet, John~E Dennis~Jr, and S{\'e}bastien~Le Digabel.
\newblock ``Parallel space decomposition of the mesh adaptive direct search
  algorithm''.
\newblock \href{https://dx.doi.org/10.1137/070707518}{SIAM Journal on
  Optimization {\bf 19}, 1150--1170}~(2008).

\bibitem{jax2018github}
James Bradbury, Roy Frostig, Peter Hawkins, Matthew~James Johnson, Chris Leary,
  Dougal Maclaurin, George Necula, Adam Paszke, Jake Vander{P}las, Skye
  Wanderman-{M}ilne, and Qiao Zhang.
\newblock ``{JAX}: composable transformations of {P}ython+{N}um{P}y programs''.
\newblock \url{http://github.com/jax-ml/jax}~(2018).

\bibitem{kirby2026observation}
William Kirby, Bibek Pokharel, Javier~Robledo Moreno, Kevin~C. Smith, Sergey
  Bravyi, Abhinav Deshpande, Constantinos Evangelinos, Bryce Fuller, James~R.
  Garrison, Ben Jaderberg, Caleb Johnson, Petar Jurcevic, Su~un~Lee, Simon
  Martiel, Mario Motta, Seetharami Seelam, Oles Shtanko, Kevin~J. Sung, Minh
  Tran, Vinay Tripathi, Kazuhiro Seki, Kazuya Shinjo, Han Xu, Lukas Broers,
  Tomonori Shirakawa, Seiji Yunoki, Kunal Sharma, and Antonio Mezzacapo.
\newblock ``Observation of improved accuracy over classical sparse ground-state
  solvers using a quantum computer''~(2026).
\newblock  \href{http://arxiv.org/abs/2603.03496}{arXiv:2603.03496}.

\bibitem{ferris2012perfect}
Andrew~J. Ferris and Guifre Vidal.
\newblock ``Perfect sampling with unitary tensor networks''.
\newblock \href{https://dx.doi.org/10.1103/PhysRevB.85.165146}{Phys. Rev. B
  {\bf 85}, 165146}~(2012).

\bibitem{bravyi2002fermionic}
Sergey~B. Bravyi and Alexei~Yu. Kitaev.
\newblock ``Fermionic quantum computation''.
\newblock \href{https://dx.doi.org/10.1006/aphy.2002.6254}{Annals of Physics
  {\bf 298}, 210--226}~(2002).

\bibitem{sharma2017semistochastic}
Sandeep Sharma, Adam~A. Holmes, Guillaume Jeanmairet, Ali Alavi, and C.~J.
  Umrigar.
\newblock ``Semistochastic heat-bath configuration interaction method: Selected
  configuration interaction with semistochastic perturbation theory''.
\newblock \href{https://dx.doi.org/10.1021/acs.jctc.6b01028}{Journal of
  Chemical Theory and Computation {\bf 13}, 1595--1604}~(2017).

\bibitem{holmes2016heat}
Adam~A. Holmes, Norm~M. Tubman, and C.~J. Umrigar.
\newblock ``Heat-bath configuration interaction: An efficient selected
  configuration interaction algorithm inspired by heat-bath sampling''.
\newblock \href{https://dx.doi.org/10.1021/acs.jctc.6b00407}{Journal of
  Chemical Theory and Computation {\bf 12}, 3674--3680}~(2016).

\bibitem{cato2013exploring}
Michael~A. Cato, Jr., D.~Majumdar, Szczepan Roszak, and Jerzy Leszczynski.
\newblock ``Exploring relative thermodynamic stabilities of formic acid and
  formamide dimers – role of low-frequency hydrogen-bond vibrations''.
\newblock \href{https://dx.doi.org/10.1021/ct300889b}{Journal of Chemical
  Theory and Computation {\bf 9}, 1016--1026}~(2013).

\bibitem{frey2006ab}
Jann~A. Frey and Samuel Leutwyler.
\newblock ``An ab initio benchmark study of hydrogen bonded formamide dimers''.
\newblock \href{https://dx.doi.org/10.1021/jp064730q}{The Journal of Physical
  Chemistry A {\bf 110}, 12512--12518}~(2006).

\bibitem{sun2018pyscf2}
Qiming Sun, Timothy~C. Berkelbach, Nick~S. Blunt, George~H. Booth, Sheng Guo,
  Zhendong Li, Junzi Liu, James~D. McClain, Elvira~R. Sayfutyarova, Sandeep
  Sharma, Sebastian Wouters, and Garnet Kin-Lic Chan.
\newblock ``{PySCF}: the {Python}-based simulations of chemistry framework''.
\newblock \href{https://dx.doi.org/10.1002/wcms.1340}{WIREs Computational
  Molecular Science {\bf 8}, e1340}~(2018).

\bibitem{sun2020pyscf}
Qiming Sun, Xing Zhang, Samragni Banerjee, Peng Bao, Marc Barbry, Nick~S.
  Blunt, Nikolay~A. Bogdanov, George~H. Booth, Jia Chen, Zhi-Hao Cui, Janus~J.
  Eriksen, Yang Gao, Sheng Guo, Jan Hermann, Matthew~R. Hermes, Kevin Koh,
  Peter Koval, Susi Lehtola, Zhendong Li, Junzi Liu, Narbe Mardirossian,
  James~D. McClain, Mario Motta, Bastien Mussard, Hung~Q. Pham, Artem Pulkin,
  Wirawan Purwanto, Paul~J. Robinson, Enrico Ronca, Elvira~R. Sayfutyarova,
  Maximilian Scheurer, Henry~F. Schurkus, James E.~T. Smith, Chong Sun,
  Shi-Ning Sun, Shiv Upadhyay, Lucas~K. Wagner, Xiao Wang, Alec White,
  James~Daniel Whitfield, Mark~J. Williamson, Sebastian Wouters, Jun Yang,
  Jason~M. Yu, Tianyu Zhu, Timothy~C. Berkelbach, Sandeep Sharma, Alexander~Yu.
  Sokolov, and Garnet Kin-Lic Chan.
\newblock ``Recent developments in the {PySCF} program package''.
\newblock \href{https://dx.doi.org/10.1063/5.0006074}{The Journal of Chemical
  Physics {\bf 153}, 024109}~(2020).

\bibitem{li2017github}
{Zhendong Li and Garnet Kin-Lic Chan}.
\newblock ``{Active-space-model-for-Iron-Sulfur-Clusters}''.
\newblock
  \url{https://github.com/zhendongli2008/Active-space-model-for-Iron-Sulfur-Clusters}.

\bibitem{li2017spinprojected}
Zhendong Li and Garnet Kin-Lic Chan.
\newblock ``Spin-projected matrix product states: Versatile tool for strongly
  correlated systems''.
\newblock \href{https://dx.doi.org/10.1021/acs.jctc.7b00270}{Journal of
  Chemical Theory and Computation {\bf 13}, 2681--2695}~(2017).

\bibitem{dicedocs}
The~Dice developers.
\newblock ``{Overview — Dice 0.1 documentation}''.
\newblock \url{https://sanshar.github.io/Dice/overview.html}.

\bibitem{2020SciPy-NMeth}
Pauli Virtanen, Ralf Gommers, Travis~E. Oliphant, Matt Haberland, Tyler Reddy,
  David Cournapeau, Evgeni Burovski, Pearu Peterson, Warren Weckesser, Jonathan
  Bright, St{\'e}fan~J. {van der Walt}, Matthew Brett, Joshua Wilson, K.~Jarrod
  Millman, Nikolay Mayorov, Andrew R.~J. Nelson, Eric Jones, Robert Kern, Eric
  Larson, C~J Carey, {\.I}lhan Polat, Yu~Feng, Eric~W. Moore, Jake
  {VanderPlas}, Denis Laxalde, Josef Perktold, Robert Cimrman, Ian Henriksen,
  E.~A. Quintero, Charles~R. Harris, Anne~M. Archibald, Ant{\^o}nio~H. Ribeiro,
  Fabian Pedregosa, Paul {van Mulbregt}, and {SciPy 1.0 Contributors}.
\newblock ``{{SciPy} 1.0: Fundamental Algorithms for Scientific Computing in
  Python}''.
\newblock \href{https://dx.doi.org/10.1038/s41592-019-0686-2}{Nature Methods
  {\bf 17}, 261--272}~(2020).

\bibitem{qiskit-addon-sqd}
Abdullah~Ash Saki, Stefano Barison, Bryce Fuller, James~R. Garrison,
  Jennifer~R. Glick, Caleb Johnson, Antonio Mezzacapo, Javier Robledo-Moreno,
  Max Rossmannek, Paul Schweigert, Iskandar Sitdikov, and Kevin~J. Sung.
\newblock ``{Qiskit addon: sample-based quantum diagonalization}''.
\newblock \url{https://github.com/Qiskit/qiskit-addon-sqd}.

\bibitem{gray2018quimb}
Johnnie Gray.
\newblock ``quimb: a python library for quantum information and many-body
  calculations''.
\newblock \href{https://dx.doi.org/10.21105/joss.00819}{Journal of Open Source
  Software {\bf 3}, 819}~(2018).

\bibitem{liu2025samplingerrormitigation}
Kecheng Liu and Zhenyu Cai.
\newblock ``Quantum error mitigation for sampling algorithms''~(2025).
\newblock  \href{http://arxiv.org/abs/2502.11285}{arXiv:2502.11285}.

\bibitem{temme2017errormitigation}
Kristan Temme, Sergey Bravyi, and Jay~M. Gambetta.
\newblock ``Error mitigation for short-depth quantum circuits''.
\newblock \href{https://dx.doi.org/10.1103/PhysRevLett.119.180509}{Phys. Rev.
  Lett. {\bf 119}, 180509}~(2017).

\end{thebibliography}

\end{document}